\documentclass[12pt,preprint2]{aastex631}
\newcommand{\Hb}{H$\beta$}

\newcommand{\Htwo}{H\textsubscript{2}}

\newcommand{\kms}{km~s{$^{-1}$}}

\newcommand{\cmq}{cm$^{-3}$}

\newcommand{\hr}{the Huygens Region}

\newcommand{\tC}{{$\rm \theta^1$~Ori~C}}

\newcommand{\tX}{{$\rm \theta^2$~Ori~A}}
\newcommand{\hii}{H II}
\newcommand{\hi}{H{\sc i}}

\newcommand{\oiii}{[O III]}

\newcommand{\Cii}{[C II]}

\newcommand{\oi}{[O I]}

\newcommand{\nii}{[N II]}
\newcommand{\sii}{[S II]}

\newcommand{\Caii}{Ca II}
\newcommand{\Nai}{Na I}

\newcommand{\siiia}{S III}

\newcommand{\piii}{P~III}

\begin{document}

\title{Large Scale Wind Driven Structures in the Orion Nebula}

\author{C. R. O'Dell\affil{1}}
\affil{Department of Physics and Astronomy, Vanderbilt University, Nashville, TN 37235-1807}

\author{N. P. Abel\affil{2}}
\affil{MCGP Department, University of Cincinnati, Clermont College, Batavia, OH, 45103}

\begin{abstract}

A study of \Cii\ 158 \micron\ and \hi\ 21-cm spectroscopic images plus high velocity resolution optical and ultraviolet spectra has shown the structure of the Orion Nebula to be different from that found from the study of those data separately. 
The \Cii\ features recently identified as the \Cii\ Shell is shown to be part of the Veil-B \hi\ foreground layer. Jointly called the Outer Shell, it covers the bright Huygens Region and the Extended Orion Nebula. Its maximum expansion velocity is 15 \kms. %It is seen in all our data sets. 
Closer to \tC\ there is a second
expanding shell, called the Inner Shell. It has an expansion velocity of 27 \kms\ and probably results from a more recent period of strong wind from one or more of the Trapezium stars.
% It too is seen in all our data sets. 
Even closer to  \tC\ there is a central high ionization bubble, freely expanding towards the observer but slowed in the opposite direction by photo-ionized gas coming off the Main Ionization Front.
Utilization of spectroscopic measures of the equivalent width of \Hb\ shows that the enhanced emission in \Cii\ seen just outside the visual wavelength boundaries of the Orion Nebula is not caused by limb-brightening of the Outer Shell.
This enhanced emission is due to the radiation field of the Trapezium stars being filtered by intervening residual neutral hydrogen. 
%it is due to illumination by Trapezium stars that have had most of their EUV filtered out by intervening residual neutral hydrogen.
 A velocity component near 30 \kms\ (Heliocentric) first seen in \hi\ is also present in \Cii\ and may result from a foreground cloud of the ISM. 

\end{abstract}


\section{INTRODUCTION}
\label{sec:intro}

We have used a remarkable set of \Cii\ 158 \micron\ images \citep{pabst19,pabst20} and an  \hi\ 21-cm image \citep{vdw13} to study in more detail the 3-D structure along the line of sight to the Orion Nebula (M42, NGC1976). The Pabst papers used the \Cii\ data to argue  that there is a hemispherical shell covering the Extended Orion Nebula (EON, \citep{gud08}) on the observer's side of the Orion Molecular Cloud (OMC). They presented a simple model for this expanding shell, arguing that it is driven by the stellar wind from the hot dominant star \tC\ \citep{ode17}. Although this star dominates the hydrogen ionizing radiation field, it is but one member of the rich (3500 stars \citep{hil98}) Orion Nebula Cluster(ONC). We believe that use of available \hi\ 21-cm data \citep{vdw13} complements the \Cii\ data and allows a more complete determination of the real structure.

There is a rich literature describing this nearest star formation region that contains early O spectral type stars, even though the brightest part of the Orion Nebula (the Huygens Region) and its associated EON
are smaller than \hii\ regions that dominate the literature. Its proximity and brightness means that it is the best region for studying how hot stars shape the environment of a rich young star cluster. Many proto-planetary disks are known to be associated with lower mass stars in the cluster \citep{ode96,mjm23,tom23} and the study of these proplyds illuminates the process of planet formation.

The following paper first presents the basic model of the Orion Nebula (Section~\ref{sec:basic}), then describes the data sets used (Section~ \ref{sec:data}), the locations of the samples we have employed (Section~\ref{sec:locations}), illustrations of the \Cii\ and \hi\ spectra (Section~\ref{sec:illustration}) and the results from measuring the spectra (Section~\ref{sec:results}).
The first of these results are for five profiles cutting across \hr\ and the EON (Sections~\ref{sec:northern}, \ref{sec:NS}, and \ref{sec:ThreeProfiles}). A comparison of \Cii\ and \hi\ results are then given (Section~\ref{sec:CandH}). In Section~\ref{sec:LargeS} we give results for large samples within \hr\ and the EON, with emphasis on \Cii\ results for the OMC-PDR at the interface of the ionized nebula and the OMC. A comparison of the velocity components inside \hr\ and the EON with samples in the Outer Border  (Section~\ref{sec:Outside}) is followed with the Discussion (Section~\ref{sec:Disc}) and a summary of our results (Section~\ref{sec:end}).

Although recent literature from the far-infrared and radio use Local Standard of Rest (LSR) velocities, in this paper we express all velocities as Heliocentric velocities,
appropriate for the study of one object and where there are many decades of ground-based observation results in the Heliocentric system.
Conversion from published data in LSR to Heliocentric was done by adding 18.1 \kms .

\section{BASIC MODEL FOR THE ORION NEBULA}
\label{sec:basic}
It is commonly accepted that the Huygens Region is a concave blister of well mapped photo-ionized gas on the observer's side of the OMC \citep{wen93,ode10}. In its foreground are multiple layers of features that include layers of neutral hydrogen 
\citep{vdw89,vdw90} seen in \hi\ emission and absorption, a low ionization layer (the nearer ionized layer or NIL) seen in low ionization optical emission lines \citep{gar07,ode20} and absorption lines in the Trapezium stars and \tX . More recently, emission from blue-shifted \Cii\ has been detected \citep{pabst19,pabst20} and this emission extends across much of the EON and its boundaries,  The photo-ionized gas is dominated by \tC\ in the inner Huygens Region but \tX\ becomes important in the southwest \citep{ode17}. Scattered light from the Trapezium stars and \hr\ becomes progressively more important than local emission with increasing distance from the Trapezium \citep{ode10}. 

\Cii\ observations have revealed new features in the Huygens Region and the EON, which is not surprising since that emission is driven by FUV radiation near the surface of the photon dominated region (PDR) that underlies the ionized hydrogen boundary. The main ionization front (MIF) is driven by higher energy EUV radiation with any emission from ionized carbon in the ionized blister being quenched at densities greater than 50 \cmq . We will demonstrate in this paper that other layers of \Cii\ emission lie along the line-of-sight.

We make extensive use of \hi\ hydrogen emission and absorption that arise in the foreground layers of material \citet{vdw89,vdw90,vdw13} that is collectively called Orion's Veil. Visual window \citep{ode93} and ultraviolet absorption line spectra reveal more about this foreground material. This foreground material includes a photo-ionized layer of material that has been studied through spectroscopy and modeling \citet{abel06,abel16,abel19}. 

This study employs published CO emission from deep within the PDR on the surface of the OMC, \Cii\ emission that originates from very low ionization regions in the PDR and the foreground, \hi\ absorption and emission lines along the line of sight, and \nii\ emission that must arise in the photo-ionized regions near an ionization front. 

Unfortunately, in the Orion Nebula literature different names have been used for the same features within the Orion Nebula. The Huygens Region is the optically brightest region, sharply bounded on the SE by the bright NE-SW Orion Bar. Within that region is a grouping of stars historically called the Trapezium. At 65\arcsec\ to the ESE is the young-star-rich Orion-S molecular cloud. The EON has a sharply defined bounding feature called the Rim \citet{ode10}, that extends from NE of the Trapezium counter-clockwise to the SSE. In the portion to the SE it runs nearly N-S and in \citet{ode09} is called the North-South Rim.  There is an apparent boundary continuing counter-clockwise from the SSE, ending with the NW Rim, that is more ill-defined than the N-S Rim. There is emission and scattered light beyond these boundaries of the EON, which we now designate as the Outer Border.

\section{DATA EMPLOYED}
\label{sec:data}

We have primarily employed data from two wavelength ranges, both chosen to illuminate different, but related physical features in the Orion Nebula.

\subsection{\Cii\ Spectra}
\label{sec:CiiSpec}
The \Cii\ 158 $\mu$m spectra come from a map of the entire EON \citep{pabst19,pabst20} at an angular resolution of 16\arcsec\ and re-binned to 0.3 \kms. The map has pixel size of 3\farcs525. The data were
created from observations with the upGREAT heterodyne array mounted on the SOFIA airborne observatory. A major program, these required 13 flights 
of the observatory. The data were provided directly by C. H. M. Pabst as a 3-D image, from which spectra were then formed. 

This \Cii\ spectral map was processed by \citet{kavak22a,kavak22b} into a series of east-west position-velocity cuts (PV-Cuts). \"Umit Kavak was kind enough to provide
copies of the 300 PV-Cuts at steps of 12\farcs4 covering the entire region. These were similar to those used in the study of the NW sector of the EON \citep{kavak22a} and 
apparent shocks seen in \Cii\ \citep{kavak22b}. Similar PV-Cuts are presented in \citet{pabst20}.

\subsection{\hi\ 21-cm Spectra}
\label{sec:pvdw}
We also employed \hi\ 21-cm observations covering the northern portion of the EON. Their angular resolution is 7\farcs2 $\times$ 5\farcs7 at a velocity
resolution of 0.77 \kms , and have a pixel size of 1\farcs8. They were used in a study of highly blue-shifted features in the Huygens Region \citep{vdw13} and were provided directly by Professor Paul van der Werf.
They are very complete over the Huygens Region but the strongest line-of-sight components are saturated, so that we can only record the velocities and approximate indications of their column densities. 
We use the \hi\ 21-cm  'Line' data (optical depth Tau data are also available but are inaccurate at high values of $\tau$), that normally shows absorption although there is a frequent emission-line component. When making a comparative analysis of the qualitative 21-cm \hi\ column density, we often took the inverse of the absorption profile and compared it side by side with the \Cii\ emission profile.

\section{LOCATION OF SAMPLES}
\label{sec:locations}

\subsection{The Huygens Region}
\label{sec:HRsamp}

 \begin{figure}
\epsscale{0.7}
\plotone{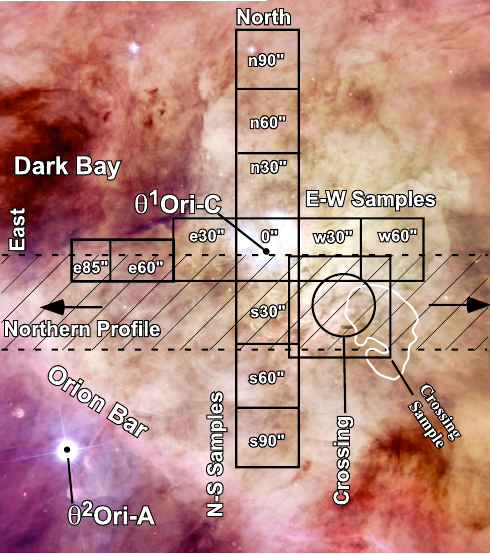}
\caption{This 235\arcsec $\times$265\farcs5 image  is extracted from a mosaic of HST images \citep{hen07}. It shows the series of samples used to determine the characteristics of the Huygens Region and the path of the samples forming the Northern Profile. The Circle is the position of the star formation complex the Crossing \citep{ode21a,ode21b,ode21c}. The white irregular outline is the \hi\ absorption  boundary of the Orion-S imbedded molecular cloud \citep{vdw13} and the surrounding box is the region sampled for analysis of its velocity components. 
\label{fig:HRsamples}}
\end{figure}

The Huygens Region is best shown in our Figure~\ref{fig:HRsamples}. This is from a mosaic of HST images created by Massimo Robberto of the Space Telescope Science Institute \citep{hen07}. 
It shows the samples discussed in Section~\ref{sec:HR}. 
Typically these samples are  30\arcsec\ $\times$ 30\arcsec\  on both the E-W and N-S directions from \tC, with the exception for the two most eastern samples that are smaller in order to avoid affects of the Dark Bay Feature.

Here and in the remainder of this paper, we frequently reference positions relative to \tC\ at 5$\rm^{h}$35$\rm^{m}$16\fs 44 -5$\arcdeg$23$\arcmin$23\farcs0. 

Fully extending across the figure is the path of a 
sequence of samples forming what we call the Northern Profile. This is centered 26\farcs2 south of \tC\ and is 38\farcs8 wide. Samples were made at intervals of 45\farcs8. The profile extends beyond this image, extending to 799\arcsec\ east and 1329\arcsec\ west of \tC.  The center line passes through the region called the Crossing \citep{ode21a,ode21b,ode21c} that is known to be the origin of multiple stellar outflows and abrupt changes in the velocity of nebular \oiii\ and \nii\ lines.

\subsection{The Northern EON}
\label{sec:2arcsec}

\begin{figure*}
\epsscale{1.0}
\plotone{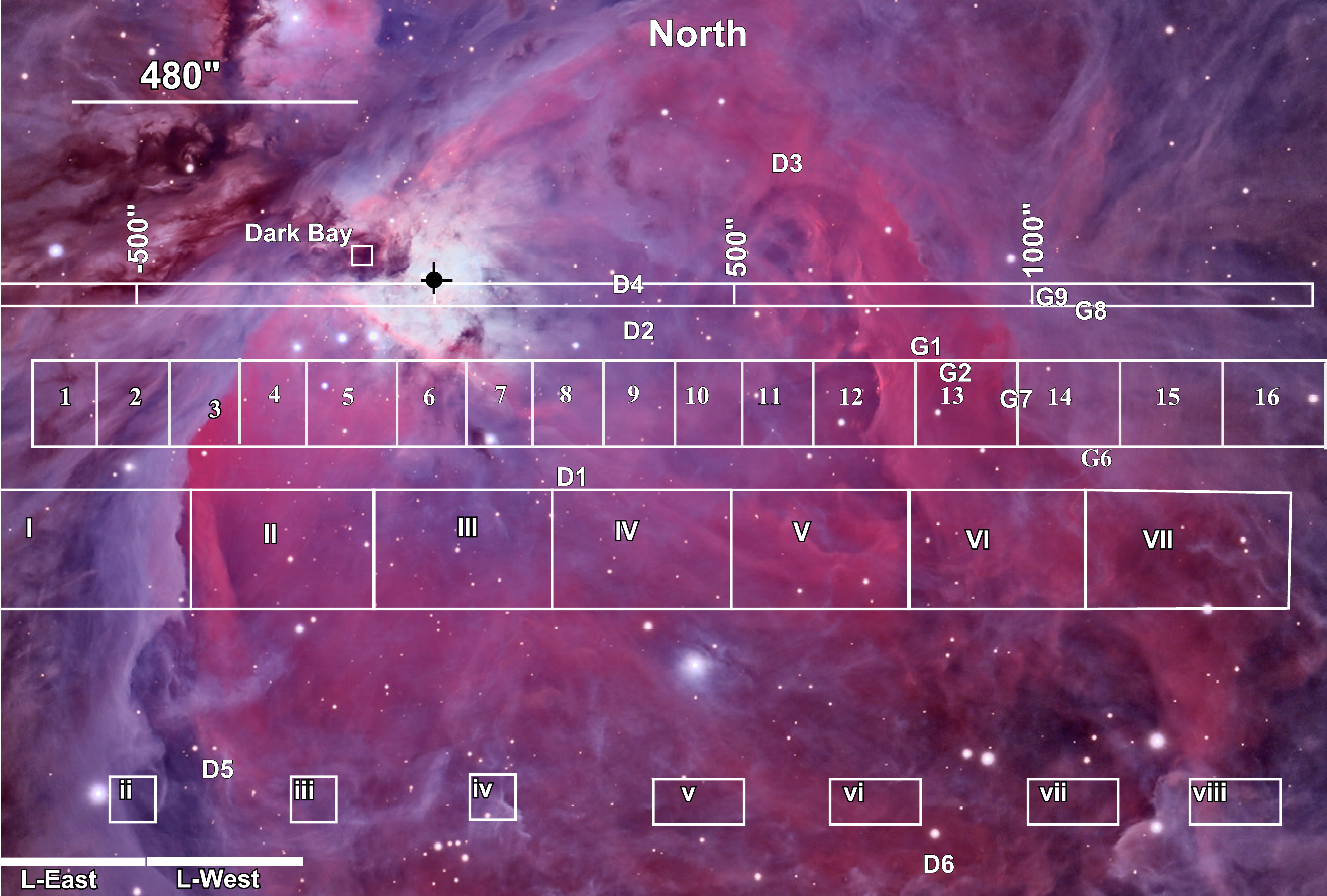}
\caption{This 2226\arcsec $\times$1514\arcsec\ image  is extracted from the Mark Manner image discussed in Appendix~\ref{sec:Mark} and shows the areas in the northern EON and the Huygens Region crossed by the path of the Northern Profile, the series of samples (labeled I-VII) that form the Middle Profile, and several of the series of samples forming the Southern Profile (labeled in lower case Roman numerals) first determined by \citet{pabst20}. 
The labeled lines (L-East, L-West) are samples overlapping the locations of optical spectra \citep{ode10}. The numbered G's indicate small areas sampled in \citet{goi20}. The filled black circle indicates the position of \tC.
\label{fig:Northernsamples}}
\end{figure*}

The Figure~\ref{fig:HRsamples} area is shown at a smaller scale in Figure~\ref{fig:Northernsamples}. In addition to the Northern Profile passing through the Crossing we show three additional cross-EON profiles, whose east-west locations are given in Table~\ref{tab:threeprofiles}.

\subsubsection{The Southern Profile}
\label{sec:southern}
The most southerly sequence are the samples of  \citet{pabst20}. They are 75\farcs5\  high, centered 880\arcsec\ south of \tC. They are designated by lower case Roman Numerals. The east-west displacements of the samples are given in Table~\ref{tab:threeprofiles}.  Although \citet{pabst20} gives the results of deconvolution into multiple \Cii\ velocity components, we have remeasured them using the same standards and methodology in the use of IRAF task 'splot' as employed in analysis of the other samples in this paper and to search for faint blue components. 

\subsubsection{The Middle Profile}
\label{sec:middle}

The next profile north of the Southern Profile is composed of ten abutting 200\arcsec\ high samples centered 450\arcsec\ south of \tC\ and are designated by upper case Roman Numerals. These cover almost all the EON along an E-W alignment and are large enough
to have high signal to noise ratios. This sequence is called the Middle Profile. It begins immediately east of the apparent east boundary of the EON and ends in the scalloped features bounding the west side of the EON. The east-west displacements of the samples are given in Table~\ref{tab:threeprofiles}.  

\subsubsection{The SE Profile}
\label{sec:SEprofile}

\begin{figure*}
\epsscale{1.5}
\plotone{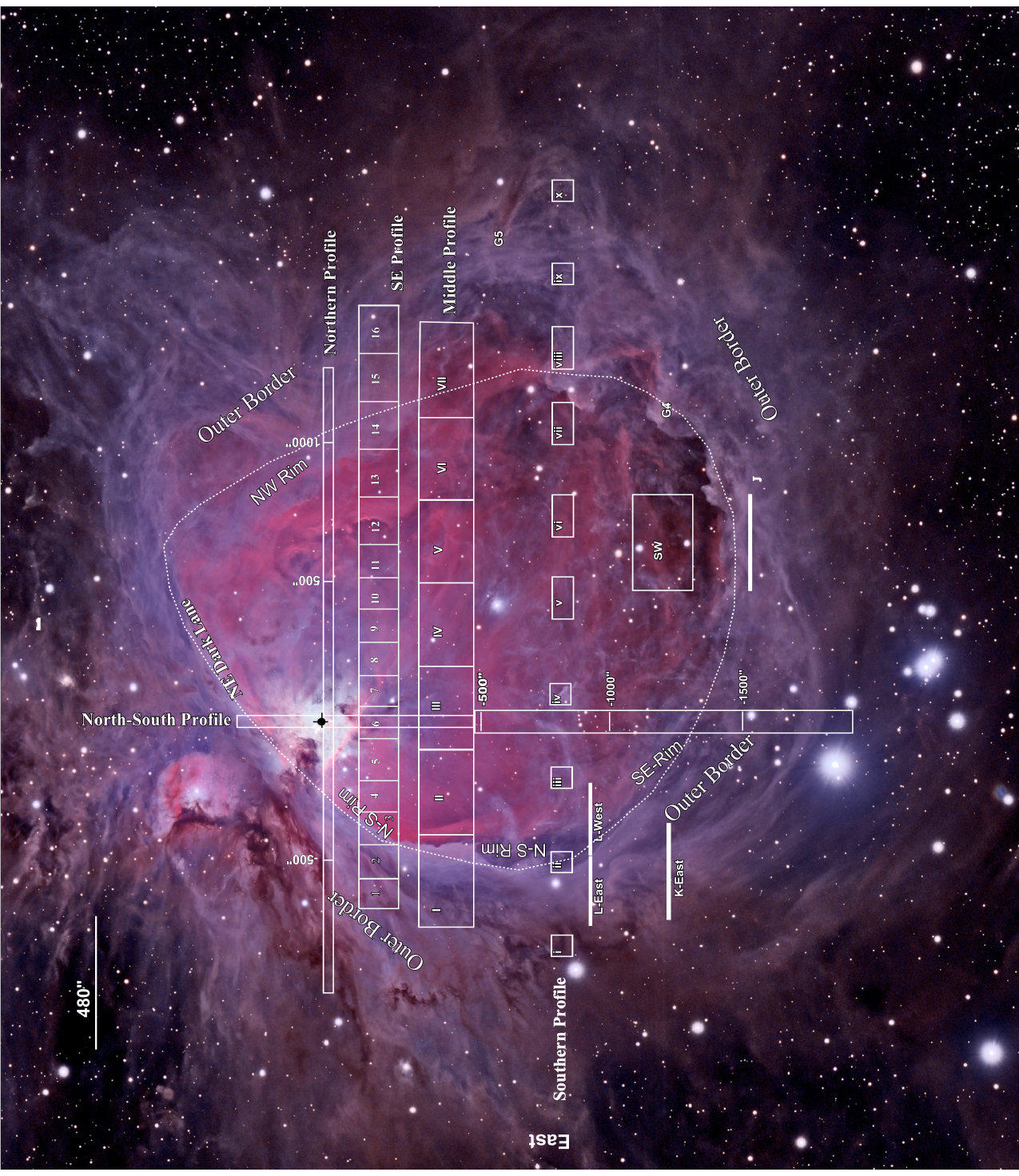}
\caption{Like Figure~\ref{fig:Northernsamples} this 4196\arcsec $\times$3694\arcsec\ image is from the from the Mark Manner image described in Appendix~\ref{sec:Mark}. In addition to most of the sample locations shown in Figure~\ref{fig:HRsamples} and Figure~\ref{fig:Northernsamples} this image adds the southernmost  samples and the location of the North-South Profile. We have designated the outermost regions as the Outer Border rather than the Limb-Brightened Region employed in previous \Cii\ studies. The white dashed line indicates what we have adopted as the boundary of \hr\ plus EON, being less well defined on the west and northwest sides. The filled black circle indicates the position of \tC. 
\label{fig:EONsamples}}
\end{figure*}

Selected to cross the EON immediately south of \hr , the SE Profile is shown in Figure~\ref{fig:Northernsamples} and Figure~\ref{fig:EONsamples}. All of the samples are 144\farcs5 high and centered 204\arcsec\ south of \tC. The east-west displacements of the samples are given in Table~\ref{tab:threeprofiles}.  

\begin{table}
\caption{\Cii\ Samples in Profiles}
\label{tab:threeprofiles}
\begin{tabular}{lll}
\hline
\hline
\colhead{SE Profile*} &
\colhead{Middle Profile**} &
\colhead{Southern Profile***}\\
\hline          
{\bf One}   (E670\arcsec --E557\arcsec)                                       &{\bf I} (E1163\arcsec --E811\arcsec) &{\bf i} (E846\arcsec --E770\arcsec) \\
{\bf Two}   (E553\arcsec --E441\arcsec)                                      &{\bf II} (E811\arcsec --E472\arcsec) &{\bf  ii}  (E544\arcsec --E468\arcsec)  \\
{\bf Three}  (E437\arcsec --E324\arcsec)                                        & {\bf III} (E472\arcsec --E100\arcsec)&{\bf iii}~(E242\arcsec --E166\arcsec) \\
{\bf Four}     (E321\arcsec --E208\arcsec)                                     & {\bf IV} (E100\arcsec --W200\arcsec)& {\bf iv} (W60\arcsec --W136\arcsec) \\
{\bf Five}    (E204\arcsec --E56\arcsec)                                      & {\bf V} (W200\arcsec --W500\arcsec)&{\bf v} (W362\arcsec --W513\arcsec) \\
{\bf Six}       (E53\arcsec --W60\arcsec)                                   &{\bf VI}  (W500\arcsec --W800\arcsec) &  {\bf vi} (W664\arcsec --W815\arcsec ) \\
{\bf Seven}  (W63\arcsec --W176\arcsec)                                        &{\bf VII} (W800\arcsec -- W1100\arcsec)&{\bf vii} (W966\arcsec --W1117\arcsec )  \\ 
{\bf Eight}    (W180\arcsec --W293\arcsec)                                      &{\bf VIII} (W1107\arcsec --W1382\arcsec)   & {\bf viii} (W1268\arcsec --W1419\arcsec )  \\
{\bf Nine}     (300\arcsec --W409\arcsec)                                     & {\bf IX} (1380W\arcsec --1657W\arcsec)         & {\bf ix} (W1570\arcsec --W1646\arcsec ) \\
{\bf Ten}      (W412\arcsec --W525\arcsec)                                    & {\bf X} (1660W\arcsec --1935W\arcsec)       &{\bf x} (W1872\arcsec --W1948 \arcsec ) \\  
{\bf Eleven}  (W529\arcsec --W642\arcsec)                                        &(...)  &(...)\\
{\bf Twelve}  (W645\arcsec --W814\arcsec)                                        &(...)  &(...)\\
{\bf Thirteen}   (W818\arcsec --W987\arcsec)                                       &(...)  &(...)\\
{\bf Fourteen} (W990\arcsec --W1160\arcsec)                                         &(...)  &(...)\\
{\bf Fifteen}   (W1163\arcsec --W1332\arcsec)                                       &(...)  &(...)\\
{\bf Sixteen}   (W1336\arcsec --W1505\arcsec)                                       &(...)  &(...)\\
\hline
\end{tabular}\\

~* All samples are centered 204\arcsec\ south of \tC.

**All Samples are centered 450\arcsec\ south of \tC

***All samples are centered 880\arcsec\ south of \tC.

\end{table}

\subsubsection{Large Samples}
\label{sec:large}

An individual large sample is shown in Figure~\ref{fig:Northernsamples}. The Dark Bay 45\arcsec\ square sample's location is centered 113\farcs5 east and 53\arcsec north of \tC .

\subsubsection{Small Samples}
\label{sec:small}

The 48\farcs5 high samples  L-East, and L-West were selected to be centered on narrow slit positions in the spectroscopic exploration of the entire Orion Nebula \citep{ode10}. 

Figure~\ref{fig:Northernsamples} also shows positions G3, G4, G5, and G10 from the study of CO features by \citet{goi20}. In their Table~A.2 they give the velocity centroids of blue-shifted \Cii\ and CO for each feature.
In addition, this figure shows the location D1-D6, indicating "dents" identified by \citet{kavak22b}. In their Appendix A they give component velocities for the dents, the Veil, and the OMC systems.

\subsection{The Full EON}
\label{sec:EON}
In Figure~\ref{fig:EONsamples} we show the full Orion Nebula, including the Huygens Region, the EON, and the Outer Border and have now added slit samples (K-East and J), and the large sample SW that is 345\arcsec $\times$ 217\arcsec\ 
 in size and centered 642\arcsec\ west and 1244\arcsec\ south of \tC .
This figure also adds the locations of the North-South Profile discussed in Section~\ref{sec:NS}. 

\section{ILLUSTRATIVE SPECTRA}
\label{sec:illustration}

\begin{figure*}           
\epsscale{1.1}
\plotone{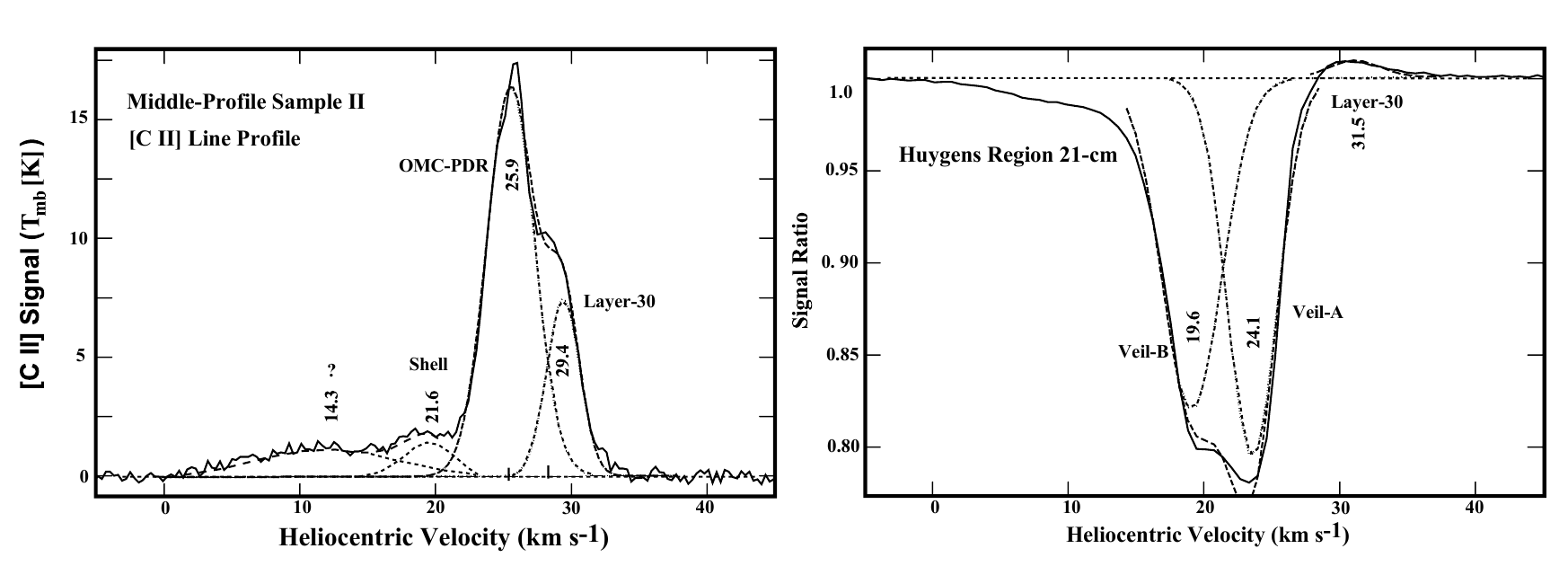}
\caption{As an illustration of the \Cii\ and 21-cm data that we used, we show in the left panel the \Cii\ line profile for Sample II of the Middle Profile, Figure~\ref{fig:Northernsamples}). The right panel shows the 21-cm spectrum for the average of the Huygens Region samples, Figure~\ref{fig:HRsamples}. 
\label{fig:CIIandLine}}
\end{figure*}

As an illustration of the spectra that we extracted from the \Cii\ and \hi\ data we show line profiles as examined with task 'splot' in Figure~\ref{fig:CIIandLine}. In all cases, we used the minimum number of components in fitting the profiles. Both the \Cii\ and \hi\ figures show blue tails, that often contain considerable energy, but defy fitting with gaussian components. 

\section{RESULTS FROM PROFILES ACROSS THE FULL ORION NEBULA}
\label{sec:results}

Although the \citet{pabst19,pabst20} studies addressed the structure of the \Cii\ emission across the entire EON using PV-Cuts, they did not explore in detail \hr , and only deconvolved  spectra in a series of samples across the middle of the EON.  In this paper we make deconvolutions at multiple cuts across the EON, supplementing these with PV-Cuts, giving more attention to the weak blue-shifted components that did not receive earlier attention. We then explore the properties of particularly
important areas within \hr.  

We first characterize the shell covering \hr\ and the entire EON (Sections~\ref{sec:northern} and \ref{sec:NS}. These in large part confirm the model presented in \citet{pabst19,pabst20}, and extend it to cover \hr . This is followed by a description of three profiles using higher S/N data from larger samples that exclusively characterize the fainter regions south of \hr. The Southern Profile 
(Section~\ref{sec:Southern} is essentially the same data as the \citet{pabst20} single profile, but extends the analysis to fainter velocity components. The Middle Profile (Section~\ref{sec:Middle} is higher in S/N due to the greater surface brightness in this area and the SE Profile (Section~\ref{sec:SE}  characterizes the region immediately south of \hr .

\subsection{The Northern Profile}
\label{sec:northern}

\begin{figure}
\epsscale{1.0}
\plotone{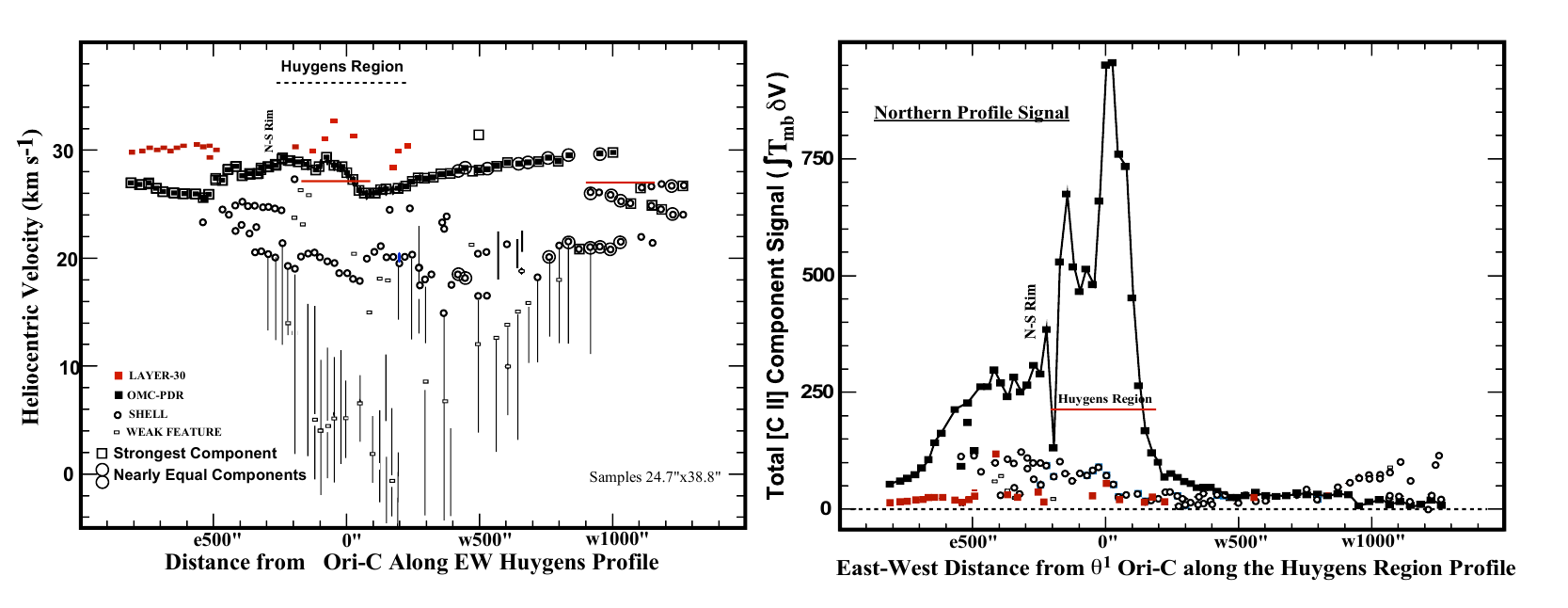}
\caption{The left panel shows the \Cii\ velocity components for positions 46\farcs8 square across the Huygens Region centered at 26\farcs2 south of \tC\ as described in Section~\ref{sec:HR}. The right panel shows the signal strength of the components in the samples whose velocities are shown in the left panel. 
\label{fig:EWProfiles}} 
\end{figure}

The Northern Profile crosses the best studied portion of the Orion Nebula (\hr ). In the left hand panel of Figure~\ref{fig:EWProfiles} we show the velocity components and in the right hand panel we show the total line signal of these components. The consistently strong OMC-PDR components have a velocity about 27 \kms , comparable to the average CO velocity of 27.3 \kms\ found by \citet{pabst24} for \hr . Multiple Layer-30 components are seen in the east end, well beyond \hr , but a few are found with \hr. Fig. A.8 in \citet{goi20} shows a PV diagram for \Cii\ and $\rm ^{12}$CO (2-1) that overlaps the Northern-Profile. 

We see that the \Cii\ OMC-PDR velocity components occur well beyond \hr , consistent with Trapezium star FUV radiation extending well beyond the \tC\ EUV that dominates \hr.

\begin{figure}
\epsscale{1.0}
\plotone{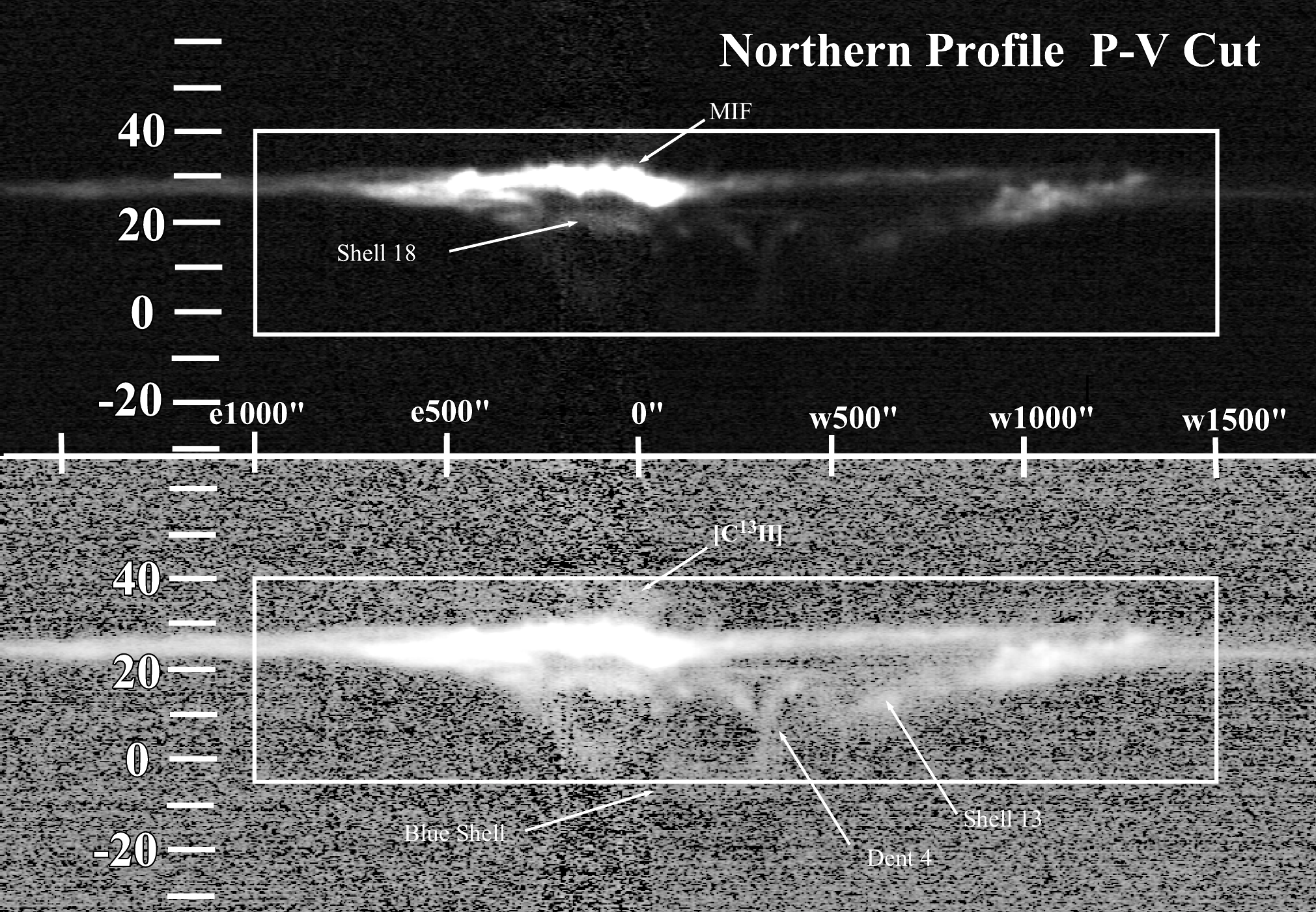}
\caption{This 3420\arcsec\ wide pair of images shows the average of three \Cii\ PV-Cuts crossing the Northern Profile samples shown in Figure~\ref{fig:Northernsamples}, whose results are shown in Figure~\ref{fig:EWProfiles}. The white box is the area covered in Figure~\ref{fig:EWProfiles}. Vertical lines depict the Heliocentric Velocity and horizontal lines the
displacements from \tC . The upper panel is a linear display, allowing the bright features along the MIF to be discerned and the lower panel is a logarithmic display of the same data, now showing the fainter more blue-shifted emission. In this and subsequent PV-Cuts the designation "Shell Number" indicates the minimum velocity of an apparent shell feature. Dent 4 is a feature discovered and discussed in \citet{kavak22b}. 
\label{fig:PVnorthern}} 
\end{figure}

Although there is a large scatter and the hint of substructure, there is evidence of a concave pattern of velocities, with a maximum separation from the OMC-PDR of about 9 \kms\ (marked as Shell 18 in Figure~\ref{fig:PVnorthern}.  A partial shell (Shell 13) appears on the west side, eventually merging with the velocities of the primary feature. 
 On the east end of the Northern Profile the Shell components approach the velocity of the OMC-PDR well outside the N-S Rim that defines the eastern edge of \hr\ \citep{ode09}. The western boundary of \hr\ is not well defined but the Shell appears to pass the velocity of the Shell 20 feature and approach the OMC-PDR emission velocity at about 1300\arcsec\ east.
 
We see in Figure~\ref{fig:EWProfiles} and Figure~\ref{fig:PVnorthern} that there is second but faint concave feature that reaches about -3 \kms, i.e. blue-shifted about 30 \kms\ from the OMC. We designate it here as the Blue Shell.  The minimum component velocities are very similar to the low ionization layer (the NIL \citep{ode20} discussed extensively using optical and UV data. There is a slight hint of an even more highly blue-shifted shell that we have labeled the High Velocity Shell.  

Although the continuous nature of the blue shifted  material had not been reported before, \Cii\ knots at these blue velocities had been noted previously \citep{kavak22b} from a search looking for compact blue shifted features.

\subsection{The North-South Profile}
\label{sec:NS}

A North-South (N-S) Profile was also created, as shown in Figure~\ref{fig:EONsamples}. The zero point is 26\farcs2 south of \tC . North of -515\arcsec\ the samples are 45\farcs8$\times$45\farcs8\ and south of there 81\farcs1$\times$74\farcs0. 

The dominant feature is the \Cii\ emission from the OMC-PDR. The open circles show the \Cii\ Shell, the dip in velocities confirming  this interpretation. The minimum value is about  16 \kms\, occurring about 800\arcsec\ south, that corresponds to a shell expansion velocity of 11 \kms .  Closure of the \Cii\ Shell at the south is likely but would occur well into the Outer Border. Closure on the north end is uncertain since
the structure of the region becomes very complex at the NE Dark Lane \citep{ode10}.

A few additional notes on this profile are in order. We note that the Layer-30 velocity component extends 500\arcsec\ south, well beyond the Huygens Region.
Disruptions of the \Cii\ Shell component near the Orion Bar again \citep{vdw13,abel16} argue for this component being closer to \tC\ than component \hi\ Veil-A.  An unusually strong red component at 30.2 \kms\ in the s1625\arcsec\ sample was isolated and identified with the binary M3.3 star 2MASSJ05351602-0550448.

\begin{figure*}
\epsscale{1.1}
\plotone{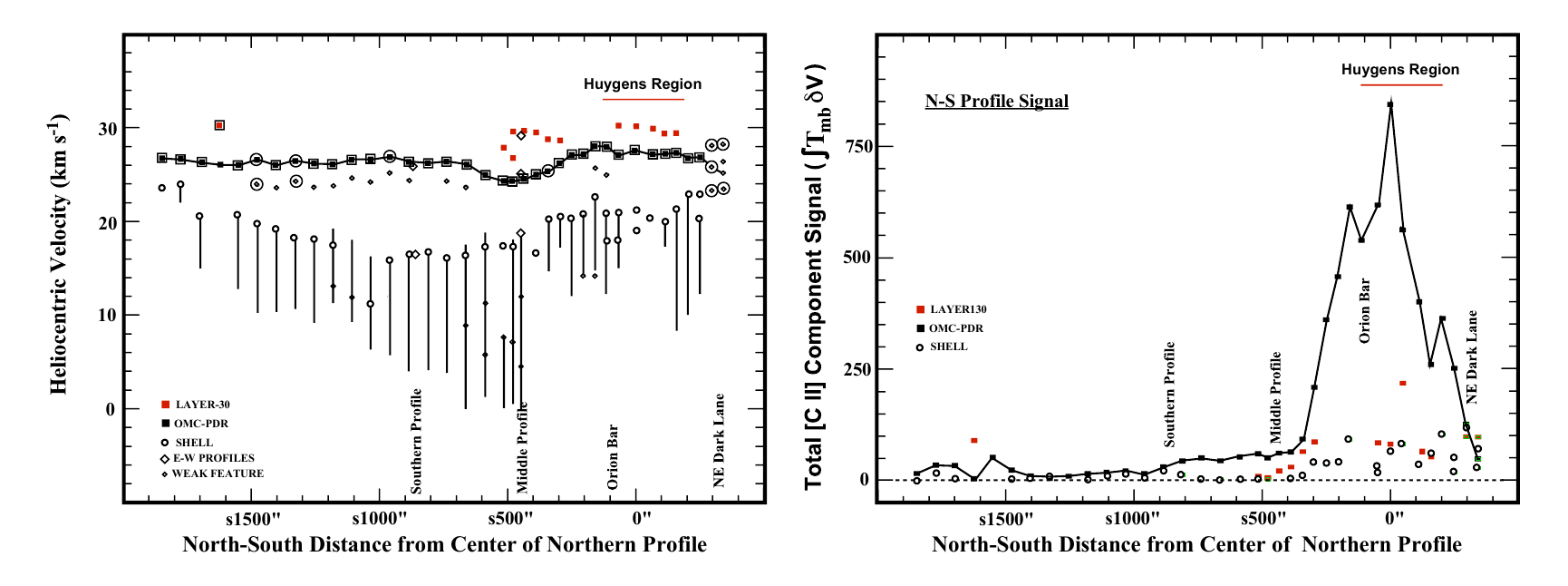}
\caption{Like Figure~\ref{fig:EWProfiles} except now showing the results for a North-South Profile. 
The large open diamonds show the results of nearby Southern and Middle Profiles samples (shown in Figures~\ref{fig:SouthernProfile} and \ref{fig:MiddleProfile} respectively). A PV-Cut of this profile is shown in \citet{pabst20} in Fig. C.2 Appendix C, labeled by its east-west displacement of 0.0. 
\label{fig:NSProfiles}}
\end{figure*}
\newpage
\subsection{Central Three Profiles}
\label{sec:ThreeProfiles}

We created three additional profiles in fainter regions of the EON. These involve larger samples, allowing better S/N ratios, and cover from the central EON north through immediately south of \hr. With one exception (SE Profile position 4) reliable \hi\ data was not available. Therefore this section is primarily a discussion of the \Cii\ data. 

\subsubsection{The Southern Profile}
\label{sec:Southern}
\begin{figure*}
\epsscale{0.7}
\plotone{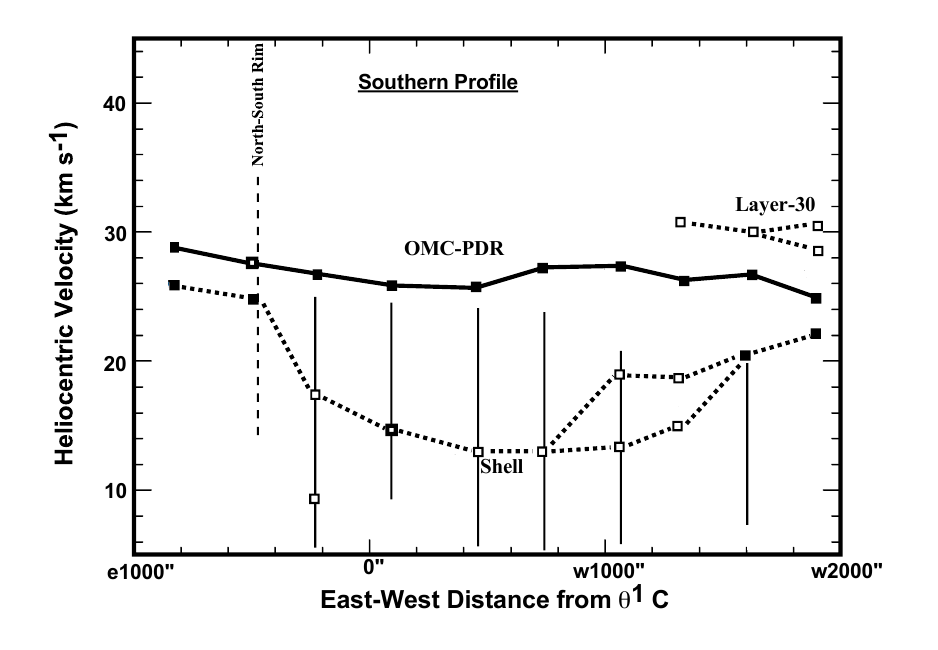}
\caption{Results for \Cii\ in the Southern Profile are shown. Filled squares indicate the strongest velocity component in that sample, heavy open squares indicate components almost as strong as the strongest component, and light-line squares indicate weaker components. Vertical lines indicate regions of continuous emission. 
The OMC-PDR sequence was identified by velocities near 27.3 \kms\ as found for CO in this region.
\label{fig:SouthernProfile}}
\end{figure*}

\begin{figure*}
\epsscale{0.9}
\plotone{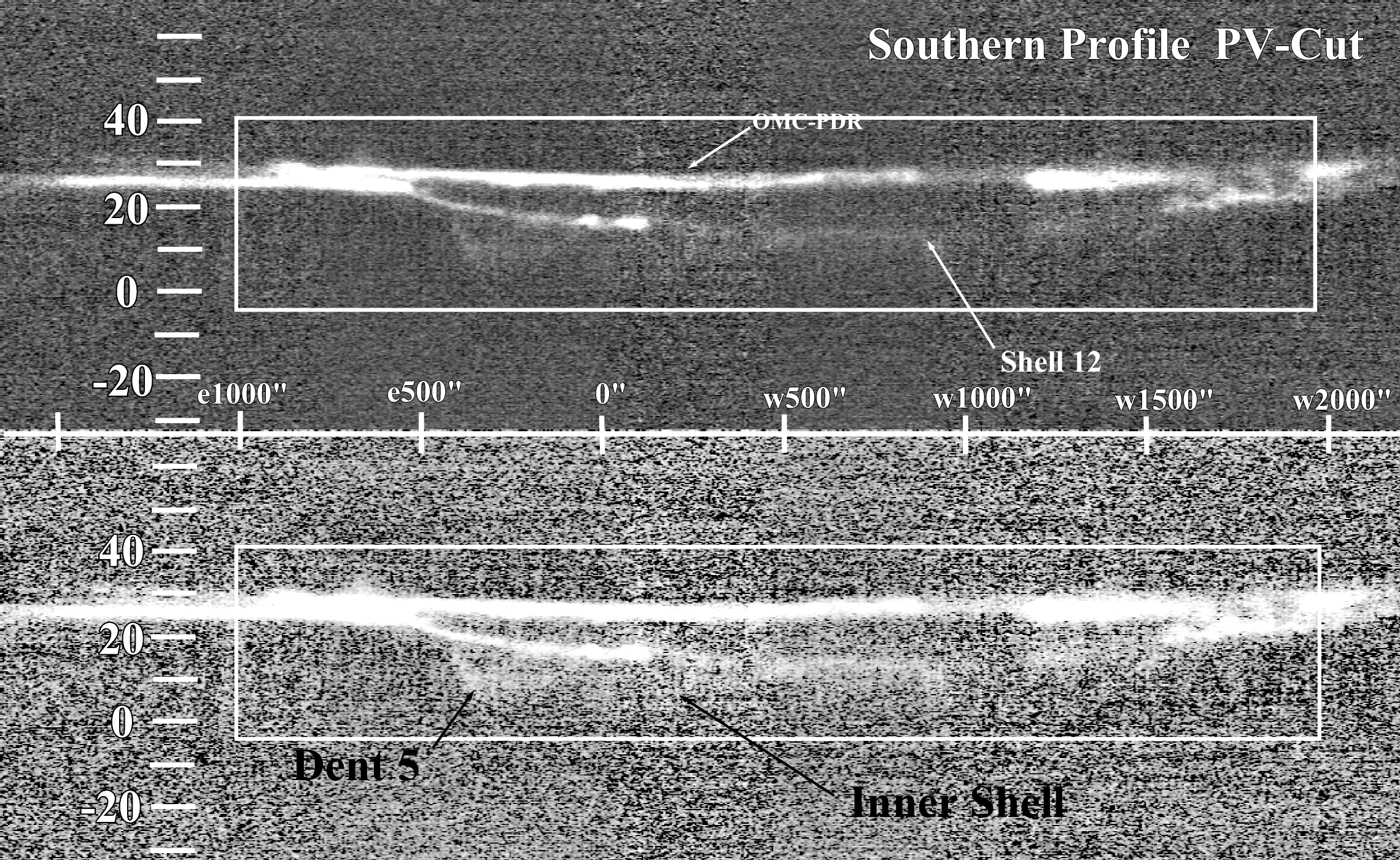}
\caption{This 3860\arcsec\ wide pair of images shows the average of five \Cii\ PV-Cuts crossing the Southern Profile samples shown in Figure~\ref{fig:Northernsamples} and whose results are shown in Figure~\ref{fig:SouthernProfile}. The white box is the area covered in Figure~\ref{fig:SouthernProfile}. Vertical lines depict the Heliocentric Velocity and horizontal lines the
displacements from \tC . Like Figure~\ref{fig:PVnorthern}, the upper panel is a linear display, allowing the bright features along the MIF to be discerned and the lower panel is a logarithmic display of the same data, now showing the fainter more blue-shifted emission. 
This is essentially a monochrome version of Figure~5 of  \citet{pabst20}.
\label{fig:PVsouthern}} 
\end{figure*}

The location and coincidence of our samples with the \citet{pabst20} study is described in Section~\ref{sec:southern}
The Southern Profile shows well the \Cii\ Shell structure  discovered in \citet{pabst19} and it has a maximum separation from the MIF component of 15 \kms , that corresponds to the expansion velocity of the shell at that declination. The quantitative Southern Profile is shown in Figure~\ref{fig:SouthernProfile} and an annotated PV diagram in Figure~\ref{fig:PVsouthern}. We see weak features bluer than the \Cii\ Shell in \hr\ that we label the Inner Shell, but it lacks the shell properties seen in the Northern Profile. 

\subsubsection{The Middle Profile}
\label{sec:Middle}

\begin{figure*}
\epsscale{0.7}
\plotone{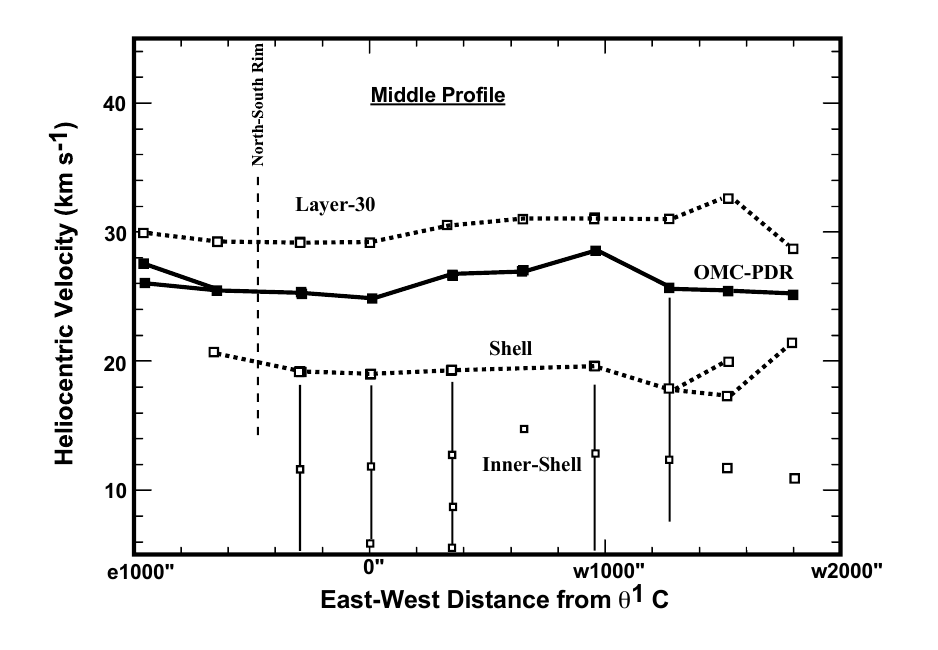}
\caption{The same symbol coding as in Figure~\ref{fig:SouthernProfile} for this \Cii\ profile. The PV diagram is shown in Figure~ \ref{fig:PVmiddle}.
\label{fig:MiddleProfile}}
\end{figure*}

The location of the Middle Profile is described in Section~\ref{sec:middle}. The Middle Profile is shown in Figure~\ref{fig:MiddleProfile}. The OMC-PDR sequence was identified by velocities near 27.3 \kms\ as found for CO in this region. 
There is little variation of velocity in the \Cii\ Shell components in the Middle Profile. There is a maximum separation from the MIF component of about 8 \kms\ and only a few changes of velocity indicate membership in the shell. One expects an  increase in velocity at the two ends of the Shell components and this is not seen. This probably indicates
that the shell is wider than in the Southern Profile. The lower velocity separation of the OMC-PDR and \Cii\ Shell components indicate that the shell's expansion velocity is
lower in this region.

\begin{figure*}
\epsscale{0.9}
\plotone{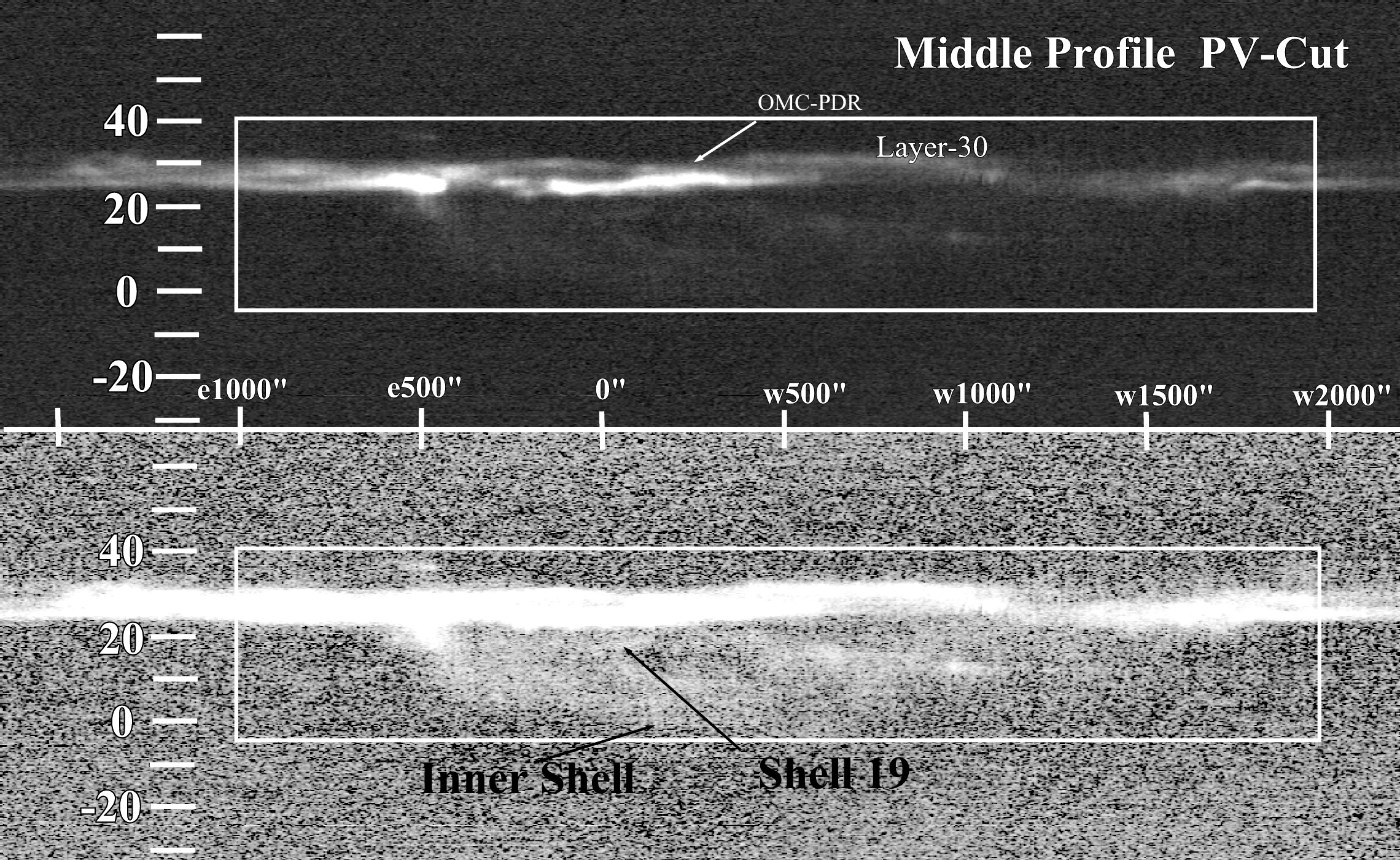}
\caption{Like Figure~\ref{fig:PVsouthern} except now showing the average for 15 PV-Cuts across the Middle Profile.
\label{fig:PVmiddle}} 
\end{figure*}

\subsubsection{The SE Profile}
\label{sec:SE}

\begin{figure*}
\epsscale{0.8}
\plotone{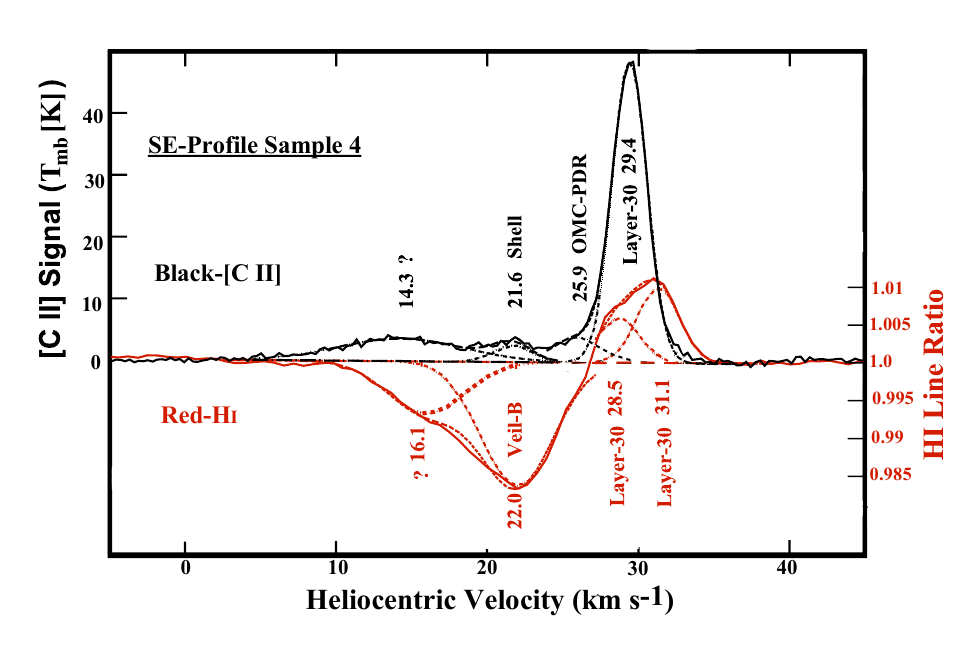}
\caption{SE-Profile Sample-4 is the first sample west of the North-South Rim of the EON boundary. The symbol coding is the same as in Figure~\ref{fig:SouthernProfile}. The \Cii\ emission-line profile is shown in black and the \hi\ line profile in red. Results for the splot deconvolution of the line profiles are shown and their assignments to velocity systems are discussed in Section~\ref{sec:SE}.
\label{fig:SE4}}
\end{figure*}

The difficulty of interpreting the SE-Profile is illustrated in Figure~\ref{fig:SE4}. This sample (c.f. Figure~\ref{fig:Northernsamples}) is in a bright region, although outside \hr , and within the EON boundary marked by the North-South Rim.  The region is bright enough that the \hi\ line ratio can be accurately measured.  The well defined and strongest \Cii\ component at 29.4 \kms\ falls within the usual range for the Layer-30 features. In contrast, the 25.9 \kms\ is closer to the 27.3 \kms\ expected for the OMC-PDR. It is not obvious why the Layer-30 component should be dominant in this sample. The region is complex as we see that the \hi\ emission features are uniquely double at 28.5 \kms\ and 31.1 \kms. The \Cii\ Shell component at 21.6 \kms\ is similar to the \hi\ Veil-B in this part of the nebula and the \Cii\ 14.3 \kms\ component is similar to the \hi\ 16.1 \kms\ component, indicating a common origin.

\begin{figure}
\epsscale{0.8}
\plotone{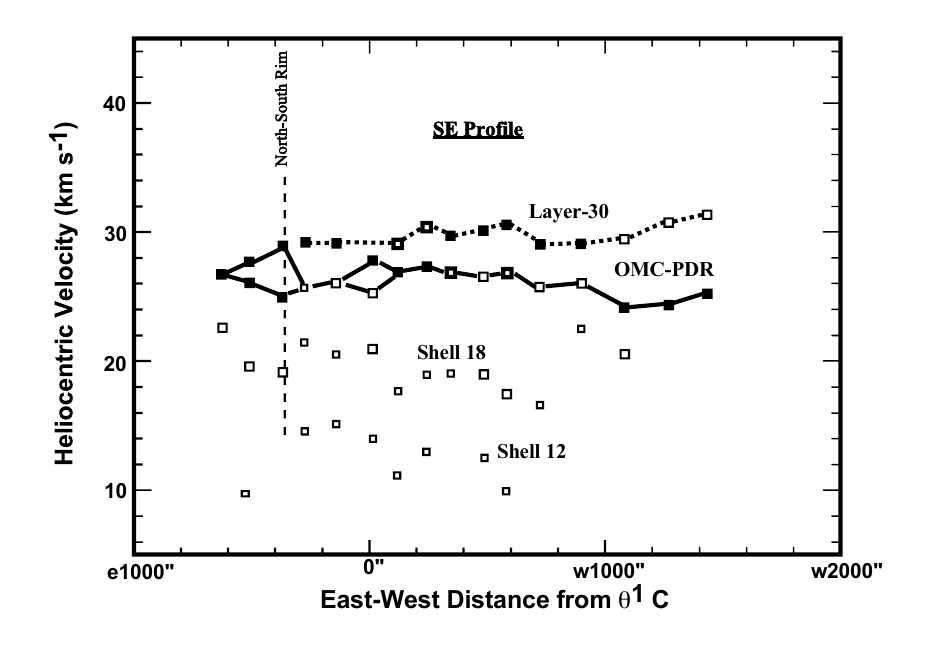}
\caption{The same figure coding as in Figure~\ref{fig:SouthernProfile} for this \Cii\ only profile. The OMC-PDR sequence components were identified as those with velocities
near those in Figure~\ref{fig:SouthernProfile} and Figure~\ref{fig:MiddleProfile}.  The PV diagram is shown in Figure~\ref{fig:PVSE}.
\label{fig:SEprofile}}
\end{figure}

The SE-Profile is shown in Figure~\ref{fig:SEprofile}.  The components connected by solid lines have been designated as the OMC-PDR. These were identified by comparing their velocities with the clearer OMC-PDR components in the Southern-Profile and the Middle-Profile. It is entirely reasonable that the higher second and third  components in the east could be part of the Layer-30 system.  It is unusual that the Layer-30 components are often the strongest or nearly the strongest components. It is notable that where (position 6)
there is a division of the OMC-PDR velocities that this sample falls on the extension of the Orion-Bar.

\newpage

\begin{figure*}
\epsscale{1.1}
\plotone{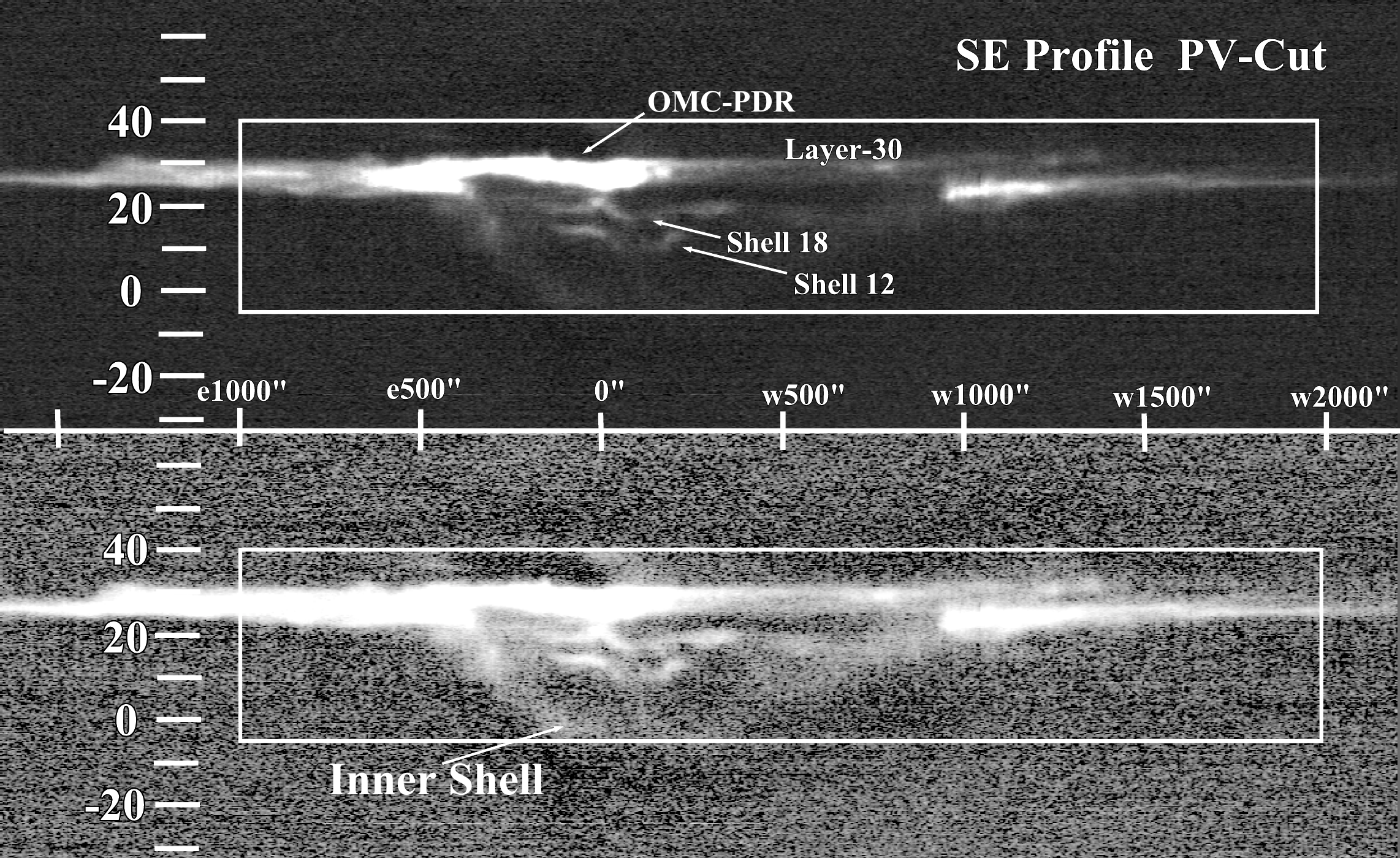}
\caption{Like Figure~\ref{fig:PVsouthern} except now showing the results for the average of nine  PV-Cuts across the SE Profile.  \label{fig:PVSE}}
 \end{figure*}
 
 \subsection{Summary of Results From the \Cii\ Profiles}
 \label{sec:summary}
 
Our multiple profiles confirm the picture of an expanding shell that includes \Cii\ emission that covers  \hr , the entire EON, and the Outer Border.
 The maximum expansion velocity is about 15 \kms\ and peaks near the center of the EON. To the north this apparent expansion velocity decreases to about 8 \kms , consistent with a radially expanding shell that in the north is viewed at an angle or a physically lower expansion velocity due to a higher overburden of ambient material. The \Cii\ Shell velocities approach those of the underlying 
 OMC-PDR in the Outer Border, but do so slowly if at all on the western side, indicating that the \Cii\ Shell may not be closed on either side.
 
 There is a previously unexplored higher blue-shifted shell of expanding gas (labeled Inner Shell) that extends down to radial velocities about 0 \kms , corresponding to an expansion velocity of 27 \kms. This Inner Shell component is highly visible
 in \hr , and decreases in visibility to the south. This decrease may be physical or represent the difficulty of detection in fainter areas.
 
  A component of emission
 at 30 \kms\ is first seen in \hr\ and the east area of the Northern Profile. The N-S profile shows that it covers \hr\ and extends south to the Middle Profile, where it is seen across the entire profile. By the Southern Profile, it is only seen in the most western samples. 

\section{COMPARISON OF \Cii\ EMISSION AND \hi\ COMPONENTS}
\label{sec:CandH}

\begin{figure*}
\epsscale{1.1}
\plotone{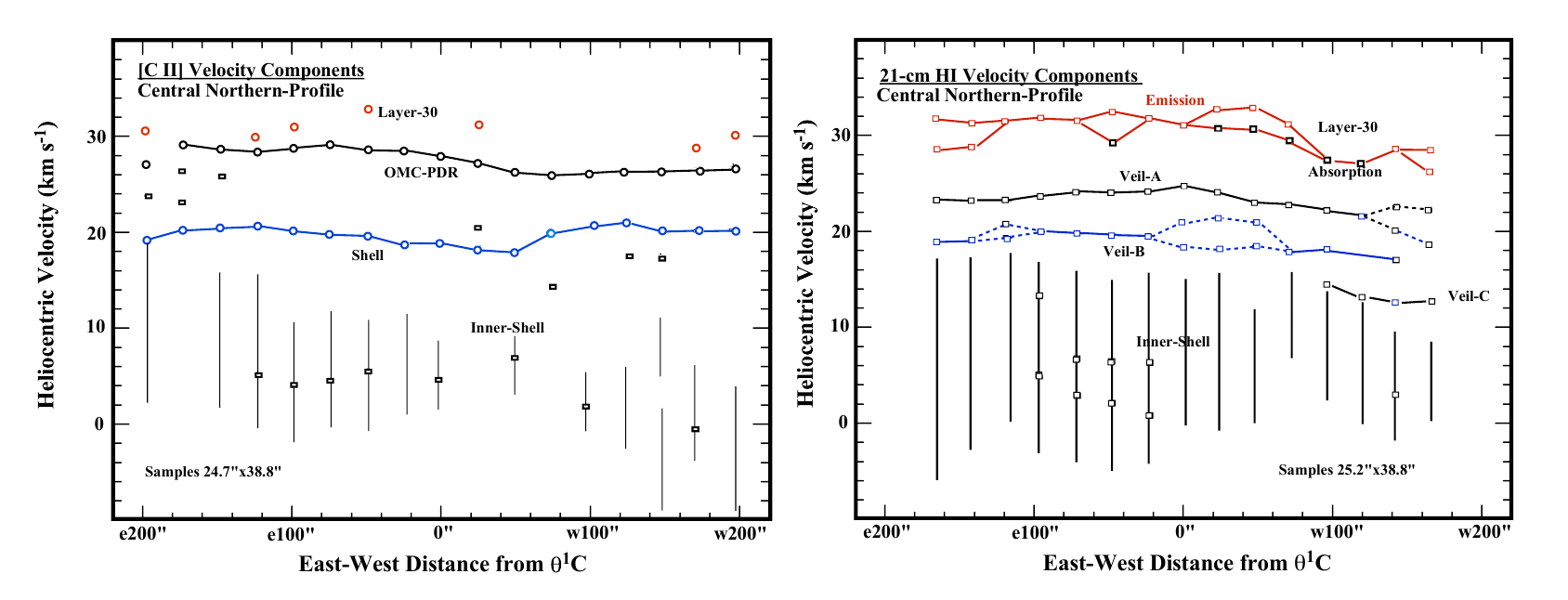}
\caption{These figures show the results of deconvolution of samples across the Central Northern Profile, now including results for \hi . Different colors and symbols are used to show common velocity sequences. In the right-hand panel, showing the \hi\ results, one sees that the weaker Veil-B  absorption appears to be split into two components. Within the \hi\ Layer-30 components, the minority that are seen in absorption are shown as black bordered boxes.
\label{fig:HReastwest}}
\end{figure*}

To examine any relation between the line-of-sight material producing the \Cii\ emission and \hi\ features, we created profiles centered on the Northern Profile but with smaller samples. The range of distances from \tC\ are limited to \hr\ since only there are the signals large enough for analysis in both. The results are shown
in Figure~\ref{fig:HReastwest}. 

The sampled regions contain the brightest areas in optical emission lines, the radio continuum, CO emission, and \Cii\ emission, and the \hi\ strongest components. It is the prime area for comparing 
all of the published information.  Although the strongest \hi\ lines are saturated, it is still possible to determine their velocities. 
\hi\ components appear both in emission and absorption while the \Cii\ components are only seen in emission. 

In the \Cii\ panel (left) one sees the strong OMC-PDR components, as expected for the blister model of the nebula.The weaker \Cii\ Shell spans the entire east to west range, but the range is small enough to not show the velocity changes characteristic of a shell. The velocity difference is 9 \kms .  At the lowest velocities we again see extended emission and discreet features (as in the Northern Profile shown in Figure~\ref{fig:EWProfiles} and collectively call them the Inner Shell.
The few Layer-30 features agree in velocity with those seen in the Huygens region in Figure~\ref{fig:EWProfiles} and Figure~\ref{fig:NSProfiles}. It is not clear that these are the only ones or if the very strong PDR emission blocks other components. 

We can discuss the \hi\ systems in the right hand panel in descending average velocity. We have adopted the assignment by  \citet{vdw89} of components near 23 \kms\ as \hi\ Veil-A and the \citet{abel19} components near 19 \kms\ as Veil-B. \citet{vdw89} also designated components near 16 \kms\ as Veil-C, later \citep{vdw13} arguing that these were part of a large-scale shock oriented to the SW.

{\bf Layer-30}  features are seen across the full span, most of the time in emission, sometimes in both emission and absorption, and in two cases in absorption only. This system was first discussed in \citet{vdw13}. The \hi\ Layer-30 velocities are essentially the same as the \Cii\ Layer-30 components, except for the region extending west from west 71\arcsec\  where the velocities drop and the absorption Layer-30 features are seen. This is the same region where optical \nii\ velocities 
have become split into two velocity systems \citep{ode21a}. Layer-30 is discussed in more detail in Section~\ref{sec:30}.

{\bf \hi\ Veil-A} at about 23 \kms\ has no counterpart in \Cii . All current studies \citep{abel19} have placed it farther from \tC\ than Veil-B. 

{\bf Veil-B} is at about 19 \kms , close to the 19.0$\pm$0.7 \kms\  \citep{abel16} most recently assigned from multiple spectroscopic evidence. This is essentially the same as the \Cii\ Shell sequence 
at about 20 \kms .The immediate question is if the \Cii\ and \hi\ sequences arise from the same physical structure. This is discussed in Section~\ref{sec:Shell-B}. It is likely that this part of the joint \Cii\ and \hi\ shell 
has the greatest initial over-burden as material is accelerated away from \tC , but will also have the greatest accelerating force, due to the intense stellar wind of \tC.  

{\bf Inner-Shell} components are again spread across the low velocities across this profile, as one sees in the \Cii\ emission.  
This component is discussed in Section~\ref{sec:BlueShells}.

In summary, we can say that the velocity data indicates that there may be links in \Cii\ and \hi\ in the Layer-30 components, the \Cii\ Shell and \hi\ Veil-B components, and the Inner-Shell components.

\section{LARGE SAMPLES FROM WITHIN THE HUYGENS REGION AND THE EON}
\label{sec:LargeS}

Examination of samples within specific areas within the EON further illuminates the local conditions and structure, rather than simply addressing the Shell. Whenever possible, this draws on both the \Cii\ and \hi\ data. Particular attention has been given to the OMC-PDR \Cii\ emission as this reflects what is happening immediately behind the MIF and we selected sample areas known to contain collimated outflows from imbedded young stars.

\subsection{The Huygens Region}
\label{sec:HR}

The Huygens Region merits full examination as it is the brightest and best studied area within the Orion Nebula, having been mapped at 10 \kms\ resolution in multiple optical emission lines \citep{gar07,gar08}, \hi\ 21-cm \citep{vdw89,vdw90,vdw13}, OH \citep{tom16} and CO \citep{pabst24} in the radio, and \Cii\ 158 $\micron$ in the mid-Infrared \citep{pabst19,pabst20}.

\begin{figure*}
\epsscale{0.8}
\plotone{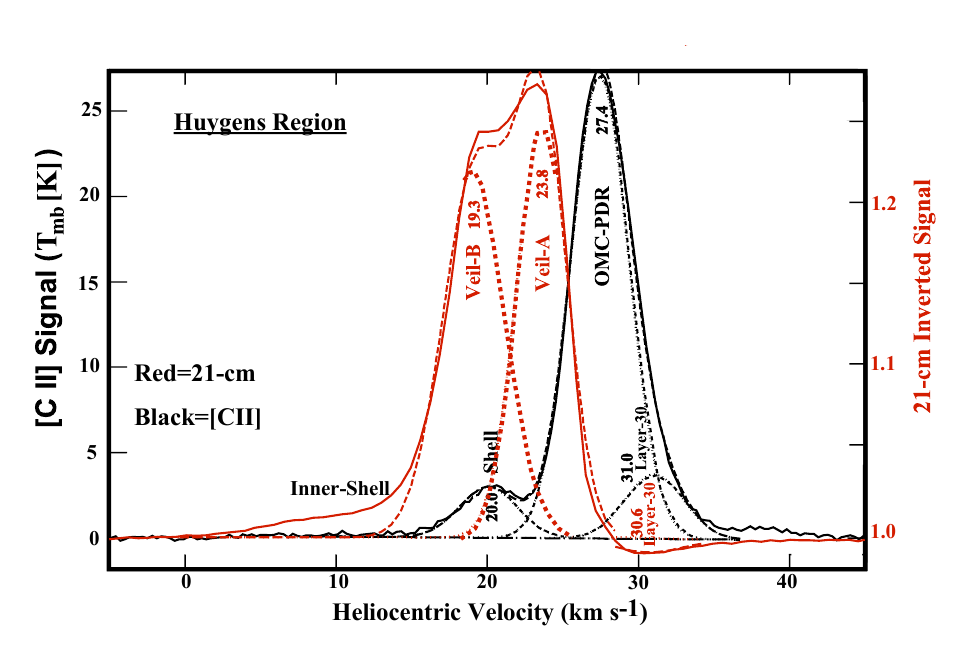}
\caption{This figure shows profiles of the average \Cii\ line superimposed on the inverted \hi\  ratio for the 12 large samples within the Huygens Region, Figure~\ref{fig:HRsamples}. 
Our samples avoided the regions of high negative velocities reported in \citet{vdw13}. 
\label{fig:HRspec}}
\end{figure*}

In Figure~\ref{fig:HRspec} we give the averaged \Cii\ and \hi\ spectra for the Huygens Region, as formed from the 12 large position labeled samples shown in Figure~\ref{fig:HRsamples}. They do not include areas where \hi\ high velocity features have been detected \citep{vdw13}. The \hi\ data is now shown as the reciprocal of the observed line profile because this aids in comparing the \Cii\ and \hi\ components. The \hi\ lines are usually seen in absorption, that appear here
above the reference line. An \hi\ dip below a value of 1.0 indicates a velocity component in emission. The results of gaussian line fits are given in the figure and in Table~\ref{tab:HRaves}. 

The velocity components are the same as those found in the profiles of \hr , but now we can see their relative signals.

Table~\ref{tab:HRaves} also gives the velocities of absorption lines for stars in \hr . We see that all components have absorption line counterparts in the stellar absorption spectra except for the OMC-PDR, placing these components nearer the observer than the bright cluster stars. 

 \newpage
\begin{table*}
\caption{Average of Velocity Systems in the Huygens Region} %and Stellar Absorption Lines$\dagger$}
\label{tab:HRaves}
\begin{tabular}{cccc}
\hline
\hline
\colhead{Assignment} &
 \colhead{V(\Cii)}&
 \colhead{V(\hi)}&
 \colhead{V(star absp line)}\\
 \hline
Layer-30          &31.0$\pm$0.8   &  30.6$\pm$0.4     & 31.0$\pm$0.3*\\
OMC-PDR        &27.4$\pm$0.5  &(...)                          &   (...)  \\
\hi\  Veil-A               &(...)                     &23.8$\pm$0.5       &22.9$\pm$0.9**\\
\hi\ Veil-B               &(...)                     &19.3$\pm$0.5       &19.2$\pm$0.7*** \\
\Cii\ Shell         &20.0$\pm$0.8        &(...)                         &(...)\\
Inner-Shell          &(...)                    \    &(...)                      &3.4$\pm$2.1****  \\
 \hline
 \end{tabular}\\
~$\dagger$All velocities are \kms\ Heliocentric. 

 ~* \Nai\ 31.3, \Caii\ 30.8 \citep{ode93}
 
 **\Nai\ 23.0, \Caii\ 22.0 \citep{ode93}, C I 23.7$\pm$0.3 \citep{abel16}.
 
 ***\Nai\ 18.3, \Caii\ 19.2 \citep{ode93}; \Htwo\ 19.6$\pm$0.8, C I 19.0$\pm$0.7 \citep{abel16}.
 
 **** He I 1.0 \citep{ode93}, \piii\ 4.8$\pm$3.0, \siiia\ 4.5$\pm$0.9 \citep{abel06}.
 
 \end{table*}

\subsection{The Crossing}
\label{sec:Crossing}
\begin{figure*}
\epsscale{1.0}
\plotone{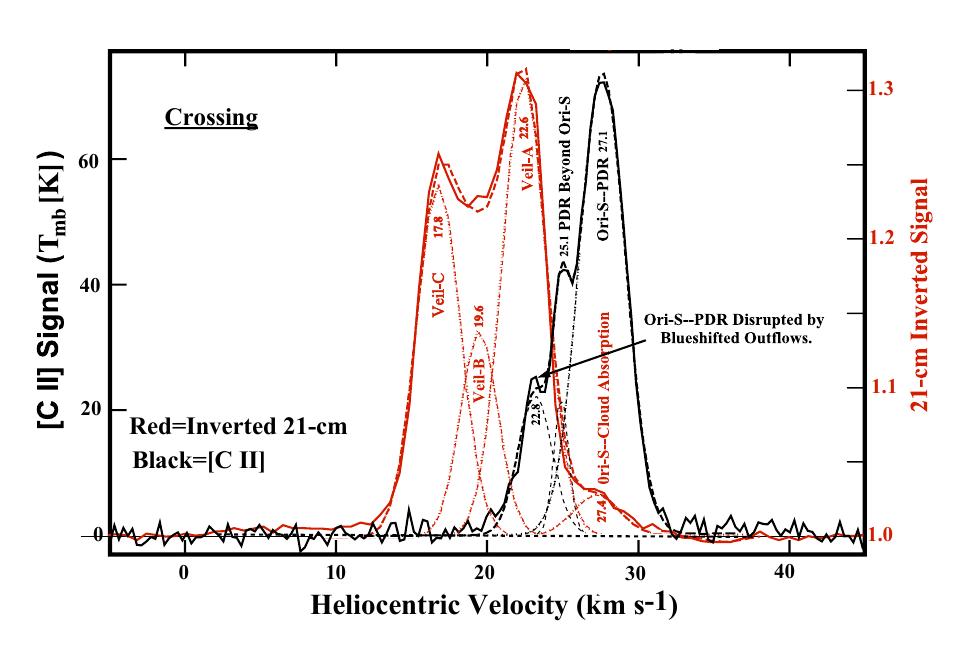}
\caption{Like Figure~\ref{fig:HRspec} except for showing the results for a large, high S/N ratio Crossing sample. The sample was composed of  52\farcs8$\times$52\farcs8 \Cii\ pixels and 45\arcsec$\times$45\arcsec\ \hi\ pixels centered 34\farcs2 west, 26\farcs2 south of \tC.
\label{fig:Crossing}}
\end{figure*}

The Crossing area is a source of multiple bipolar outflows \citep{ode15} including the large shocks HH~202, HH~203, HH~204, HH~269, and HH~529. They originate from within the embedded Ori-S molecular cloud (henceforth Ori-S or the Ori-S Cloud), then break out through the PDR on the Ori-S surface to become visible as they form shocks within the photo-ionized
lower density gas within the cavity of the cup shaped Huygens Region. Since little is known about what is happening within this cloud, we sought to use the \Cii\ and \hi\
data to explore its surface.
In Figure~\ref{fig:Crossing} we show the results for the sample shown in Figure~\ref{fig:HRsamples}, using the same presentation style as for \hr\ (Figure~\ref{fig:HRspec}).

We see only one component common to both \hr\ and the Crossing. In \Cii , \hr\ 27.4 \kms\ OMC-PDR component is almost identical with the 27.1 \kms\ in the Crossing, even though the Crossing  \Cii\ component would be formed on the observer's side of the Ori-S Cloud. The similarity of PDR velocities is explained by the internal velocity of Ori-S \citep{vdw13} being the same as the OMC. Both of these components are the brightest \Cii\ features in their regions.

An apparent agreement of the \Cii\ 22.8 \kms\ with the \hi\ Veil-A \hi\ 22.6 \kms\ is probably coincidental as the \Cii\ component would be formed in the vicinity of Ori-S, while the \hi\ Veil-A absorption component must be formed well into the foreground portions of the nebula.

The strongest \hi\ features (23.8 \kms\ in \hr, 22.6 \kms\ in the Crossing) are almost certainly due to foreground \hi\ Veil-A material.

We do not see \Cii\ and \hi\ Layer-30 features of \hr\  in the Crossing. This could be due to the embedded Ori-S Cloud lying far enough in the foreground 
to disrupt the physical feature giving rise to the Layer-30 components or, more likely, difficulties of detection. 

The strong \hi\ feature at 17.8 \kms\ in the Crossing is absent in \hr. This velocity is similar to what \citet{vdw89,vdw90} identified as a third foreground component Veil-C. In \citet{vdw13} it was argued that Veil-C was a large southwest oriented shock. We see this component in several samples to the west of the Crossing (Section~\ref{sec:HH269}, supporting the \citep{vdw13} interpretation.

Likewise, the \Cii\ Shell component seen at 20.0 \kms\ in \hr , is absent
in the Crossing, implying that the Ori-S Cloud interferes with the shell, again uncomfortably arguing that this embedded cloud is further in the foreground
than concluded from the study of visual emission-lines \cite{abel19}. It reappears to the west (Section~\ref{sec:HH269}). However, the absence may simply be blocking by the blue wing of the 22.8 \kms component.

The \hi\ 27.4 \kms\ absorption feature seen in the Crossing has no counterpart in \hr\ and must be due to absorption in the Ori-S Cloud.

The strongest \hi\ components (23.8 \kms\ in \hr , 22.6 \kms\ in the Crossing) must come from the same \hi\ Veil-A foreground material.

This leaves unexplained the unique \Cii\ components in the Crossing at 22.8 \kms\ and 25.1 \kms . 

Almost certainly there will be PDR's on the far side of the Ori-S Cloud, a region of the MIF that the observer does not see visually \citep{ode20}, but at 158 $\mu$m the Ori-S Cloud is optically thin. This is indicated by the local increase in extinction coefficient c$\rm_{H\beta}$
being only about 0.2, as shown in the study of \citet{ode00}. This means that we can see \Cii\ emission from the back-side of the Ori-S Cloud (perhaps giving the 25.1 \kms\ component) and the front side (giving the 27.1 \kms\ component). 

The \Cii\ 22.8 \kms\ and possibly the 25.1 \kms\ components may 
arise from disruption of the Ori-S--PDR as the highly blue-shifted jets driving the Herbig-Haro shocks erupt. 

\subsection{HH~269}
\label{sec:HH269}

\begin{figure*}
\epsscale{1.1}
\plotone{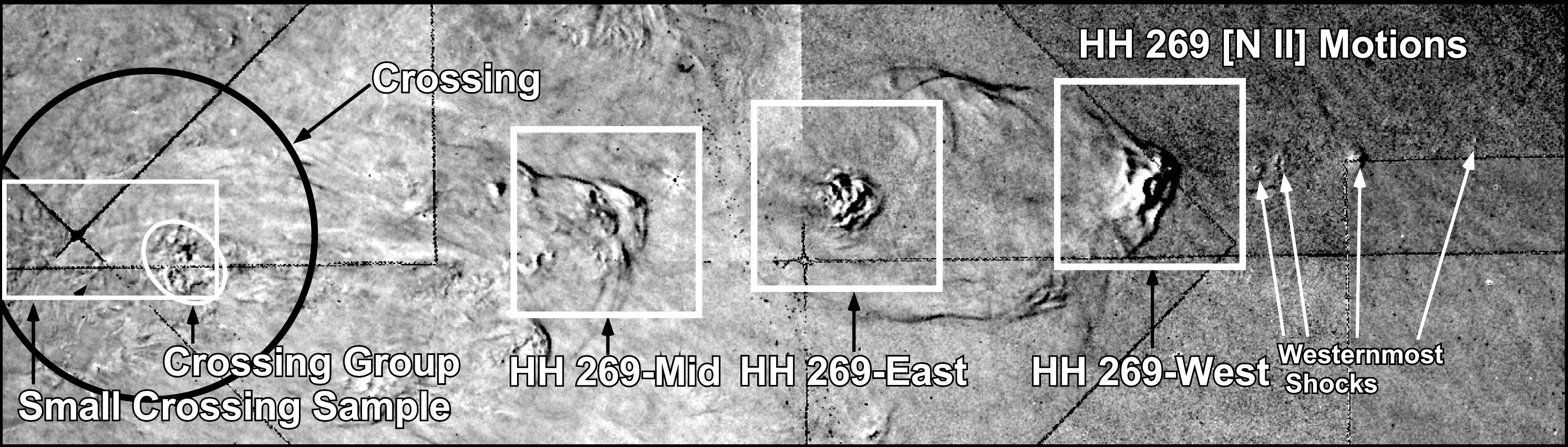}
\caption{This 143\farcs2 $\times$40\farcs6 motions image is a modified version of Figure~2 of \citet{ode21c}.  
\label{fig:HH269}}
\end{figure*}

The most extensive of the outflows from the Crossing is the  HH~269 series of shocks, whose spatial motions were discussed most recently in \citet{ode21c}.
We have sampled the major features of HH~269 in both the \hi\ and \Cii\ data.  Figure~\ref{fig:HH269} is a motions image created from ratios of HST WFPC2 F6568N
images taken over an interval of 12.6 years and reveals changed and moving structures. The dark edges indicate the direction of motion. The linear features are artifacts caused by the boundaries of the CCD detector fields of view.

We employed a series of samples. The Small Crossing sample is 10\arcsec $\times$20\arcsec\ and is centered 2\farcs5 west of the Crossing center. This sample includes the east moving emerging jet feeding HH~1132  and the complex Crossing Group. The other three samples are each 17\arcsec $\times$17\arcsec. The HH~269-Mid sample is centered 41\arcsec west and 1\farcs3 north of the Crossing center and includes the HH~269 shocks. The HH~269-East sample is 64\farcs1 west and 3\farcs6 north and contains a tight grouping of multiple shocks. The HH~269-West sample is at 91\farcs4 west and 3\farcs6 north and contains a large shock.  
The HH~269-East and HH~269-West components \citep{bom} were named before the HH~269-Mid \citep{ode15} component, producing the somewhat confusing names.

 In the \citet{vdw13} study of high velocity \hi\ (the same data used in this study) three velocity features were identified 
that fell between our samples Small Crossing and HH~269-Mid. Their component H peaked at 27.1 \kms\ and occurred atop Ori-S.  The extended Component at 22.6 \kms\ was assigned to the foreground \hi\ Veil-A feature seen in the \citet{vdw89} study. A final component they called C at 14.1 \kms\ is part of a large bow-shock structure oriented to the southwest.

\begin{figure*}
\epsscale{0.7}
\plotone{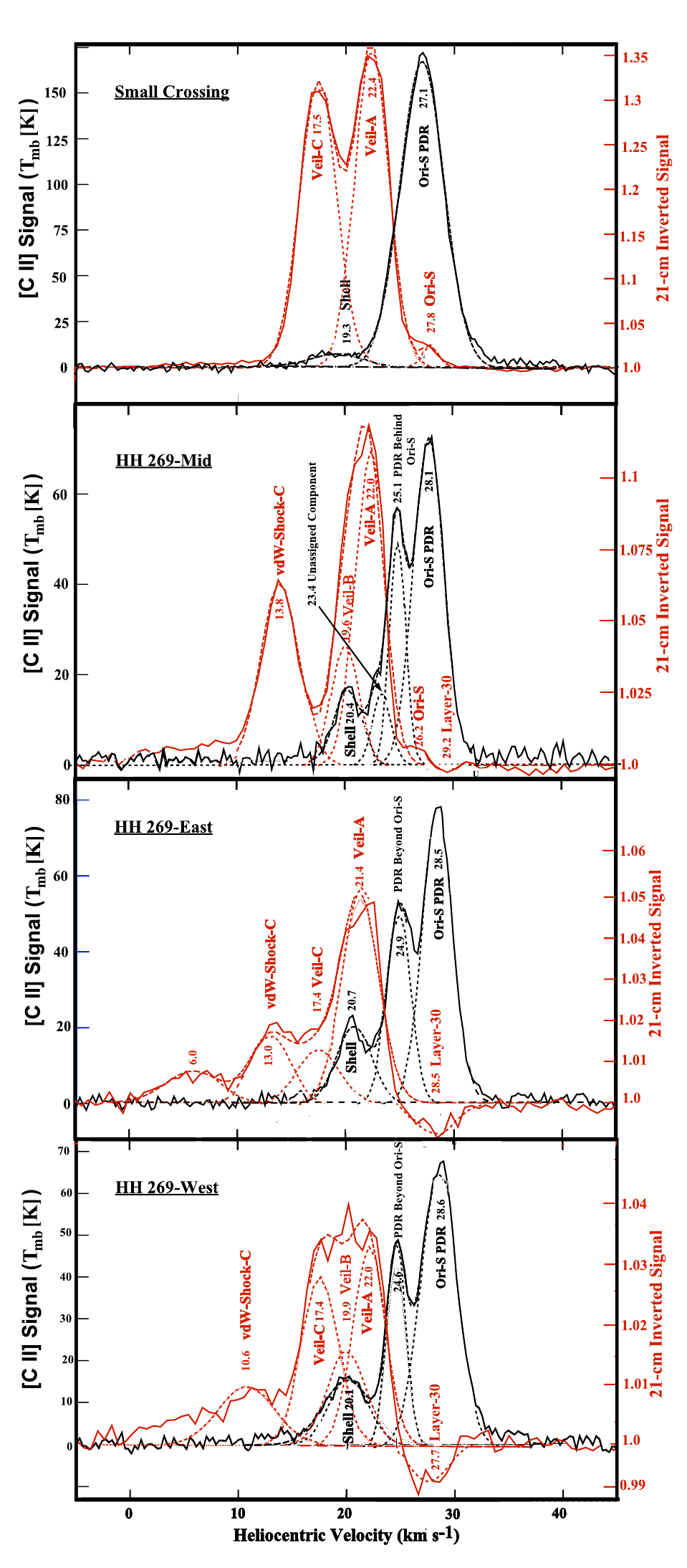}
\caption{c.f. Section \ref{sec:HH269}
\label{fig:FourPanels}}
\end{figure*}

The results for the spectra of HH~269 samples shown in Figure~\ref{fig:HH269} are given in Figure~\ref{fig:FourPanels}. The format of each panel is like that of Figure~\ref{fig:HRspec}.

{\bf \Cii\ Velocity Components:}

We identify the strong velocity components at 27.1 \kms, 28.1 \kms, 28.5 \kms\ and 28.6 \kms\  (reading from east to west) as due to the PDR of the Ori-S Cloud facing the observer, and is similar to the OMC-PDR of \hr\ at 27.4 \kms . All of these samples fall in the region lying SW of a  curved SE-NW feature caused by seeing the Ori-S photoionized material edge-on \citep{ode21a}.

As invoked in explaining the 25.1 \kms\ component in the Crossing, the HH~269-Mid (25.1 \kms), HH~269-East (24.9 \kms), and HH~269-West (24.6 \kms) probably arise from the PDR on the rear of the Ori-S Cloud, but Ori-S is transparent at long wavelengths. 

An unassigned \Cii\ component at 23.4 \kms\ is seen in HH~269-Mid. Although this is close to the \hi\  Veil-A velocity, an association is unlikely. 

A component near 20.1$\pm$0.6 \kms\ seen in each sample is nearly the same as the 20.0$\pm$0.5 \kms\ Shell component seen in \hr.

 In summary, we can say that the four samples are all covered by the Ori-S PDR, the Shell, and features behind the Ori-S Cloud. 
 
{\bf \hi\ Velocity Components:}
Things are more complex in \hi , although the \hi\ Veil-A 22.0$\pm$0.4 \kms component appears in all samples. It is, however bluer than the 23.8 \kms\ \hi\ Veil-A component in \hr. 

The reddest components must be part of Layer-30. In the HH~269-Small sample it appears in absorption at 27.8 \kms , thus being similar to the Crossing Sample at 27.4 \kms. In the more west samples (Mid 29.2 \kms , East 28.5 \kms , West 27.7 \kms ) it is seen in emission. 

A possible Veil-B component is seen in HH~269-Mid (19.6 \kms) and HH~269-West (19.9 \kms ), similar to 20.0 \kms\ in \hr .

The HH~269-Small (17.5 \kms) and HH~269-West (17.4 \kms) components are similar to the Crossing 17.8 \kms\ component and must also belong to Veil-C.

Progressing west, we see that the emission bluer than 13.8 \kms\ becomes progressively stronger, as compared with the redder \hi\ components. Only in the HH~269-East sample can it be fit to a gaussian (6.0 \kms ). Leaving the remaining samples in the category of unexplained blue emission that we see in both \Cii\ and \hi . 

 The velocity features to the blue are even more difficult to understand. Only a faint broad suggestion of bluer features is seen in HH~269-Small. \citet{vdw13} see a component at 14.1 \kms\ that appears to be part of the large southwest oriented shock and we may be seeing components of that shock. HH~268-Mid has a strong velocity component at 13.8 \kms , HH~269-East at 13.0 \kms, and HH~269-West at 
10.6 \kms. There is some \hi\ absorption near 6 \kms\ in HH~269-Mid, a well defined component at  6.0 in HH~269-East and an unresolved but real absorption at a slightly smaller velocity in HH~269-West. 

It is impossible to link the \hi\ features to the progression of \nii\ velocities of -11$\pm$3  \kms, -9$\pm$3 \kms, -13$\pm$3 \kms, and -23$\pm$2 \kms\ respectively \citep{ode21c} that mark the \nii\ shocks. Those shocks are known \citep{ode21c} to form in ionized gas lying closer to the observer than the MIF, so a correlation to \hi\ would be surprising and is not seen. The HH~269 shocks do not form where collimated jets strike foreground neutral material. This tells us that any foreground neutral layer must lie at least 0.22 pc closer to the observer than the Ori-S Cloud \citep{ode21c}. 

The \hi\ high velocity end is equally puzzling. Although in \hr\ we see a 30.6 \kms\ component in emission, there is nothing there in the HH~269-Small sample, but then there is a progression west of emission components at 29.2 \kms, 28.5 \kms, and 27.7 \kms.  If one accepts that the Small  emission velocity component is obscured by the 27.8 \kms\ absorption component, this is consistent with the entire Huygens Region being covered by a Layer of \hi\ emission at about 30 \kms.  The HH~269-Small (27.8 \kms) and HH~269-Mid (26.2) samples are closest to the Ori-S  cloud and their velocities are close to the \hi\ velocity of the cloud 27.1 \kms\ found in  \citet{vdw13}.
The ubiquity of the Layer-30 component is discussed further in Section~\ref{sec:30}.

\subsection{The Dark Bay}
\label{sec:DarkBay}

\begin{figure*}
\epsscale{0.8}
\plotone{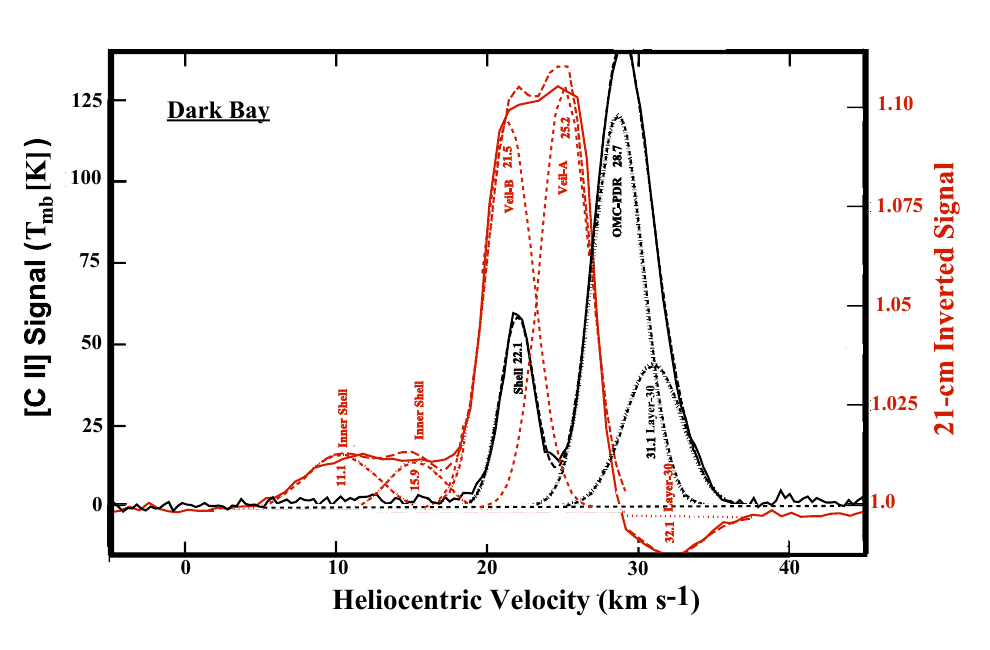}
\caption{Like Figure~\ref{fig:HRspec} except now showing the results for the Dark Bay sample. 
Common components with \hr\ are identified, although the velocities in the Dark Bay average 1.4$\pm$0.8 \kms\ larger.
\label{fig:DarkBay}}
\end{figure*}

We formed our sample of the Dark Bay to avoid the high velocity features ('E' and 'I' ) found in \citet{vdw13}. Figure~\ref{fig:Northernsamples} shows its location, that is 45\arcsec\ square and is centered 113\farcs5 east and 53\arcsec north of \tC . Most of \hr\ velocity components have apparent counterparts in \hr . However, 
the Dark Bay velocities are slightly more positive than their Huygens Region counterparts. 

{\bf \Cii} The strongest component at 28.7 \kms\ is higher than in \hr\ (27.4 \kms ) but is consistent with the rise in the OMC-PDR velocity one sees in the Northern Profile (Figure~\ref{fig:EWProfiles}. 

The 31.1 \kms\ component is consistent with other observations of Layer-30 (31.0 \kms\ in \hr).

The 22.1 \kms\ component is close to the \Cii\ Shell velocity (20.0 \kms ) in \hr .

{\bf \hi } The 32.1 \kms\ emission component agrees well with the Layer-30 component in \hr\ (30.6 \kms ), although it too is at a higher velocity than in \hr .

The strong 25.2 \kms\ absorption component is probably the analog of the 23.8 \kms\ \hi\ Veil-A component in \hr .

The strong 21.5 \kms\ absorption component is probably the analog of the 19.3 \kms\ Veil-B component in \hr. 

There are two very blue absorption components at 15.9 \kms\ and 11.1 \kms . These fall in the velocity range of the Inner-Shell \hi\ features in \hr. 
However, they also fall in the range of velocities (8 -- 14 \kms) of `Component E` \citep{vdw13} and may be associated with that feature even though our sample 
fell well outside the feature as shown in the discovery paper. 

It is remarkable that all of the components that we associate with the Orion Nebula (this excludes Layer-30) are red-shifted with respect to their Huygens Region values. The average difference is 1.9$\pm$0.4 \kms. Since this velocity difference is also true for the OMC-PDR component, this indicates 
 that the northeast section of the OMC and its foreground layers are moving redshifted by this amount. A similar result was reached in \citet{vdw13}, who noted that radial velocities increase about 0.3 \kms\ per arcminute displacement from the Trapezium
 region. Our Dark Bay sample lies 2.0\arcmin\ to the northeast and the predicted displacement would be only 0.6 \kms , in qualitative but not quantitative agreement with the earlier result. 

\subsection{The SW Sample}
\label{sec:SW}

\begin{figure*}
\epsscale{0.8}
\plotone{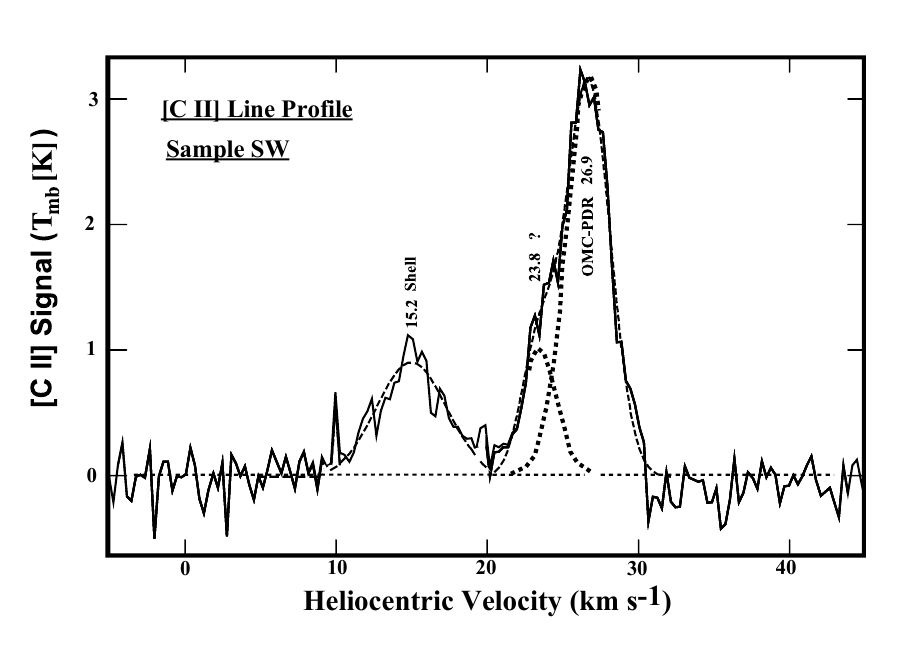}
\caption{
The \Cii\ line profile of the outlying large sample SW in the southern region of the EON is shown, as discussed in Section~\ref{sec:SW}. 
\label{fig:SW}}
\end{figure*}

In Figure~\ref{fig:SW} we show the \Cii\ line profile for a large sample that is 345\arcsec $\times$ 217\arcsec\ 
 in size and centered 642\arcsec\ west and 1244\arcsec\ south of \tC , as shown in Figure~\ref{fig:EONsamples}.
 This is the furthest EON sample from \hr\ and lies on the approximate long axis of the EON.  
 
Note the low level of the signal and that the OMC-PDR and \Cii\ Shell components dominate. The shell velocity of 15.1 \kms\ is much lower than the 26.8 \kms\ for the OMC-PDR component and is consistent with the \Cii\ Shell structure continuing in this direction. 
 
The NS-Profile shows two velocity components at about 26 \kms\ and 24 \kms\ at this distance south, with the more positive velocity component the stronger. It is likely that the 23.8 \kms\ component we see in the SW
 sample is part of this extended 24 \kms\ system. Although the velocity is similar to the \hi\ Veil-A \hi\ velocity of 23.8$\pm$0.5 \kms , an association would require FUV radiation passing through the Shell material. It is not obvious why this would occur here and not in the other regions where we don't see a \Cii\ component corresponding to \hi\ Veil-A.

\section{\Cii\ VELOCITIES FROM INSIDE AND OUTSIDE THE N-S RIM}
\label{sec:Outside}

\begin{figure*}
\epsscale{0.8}
\plotone{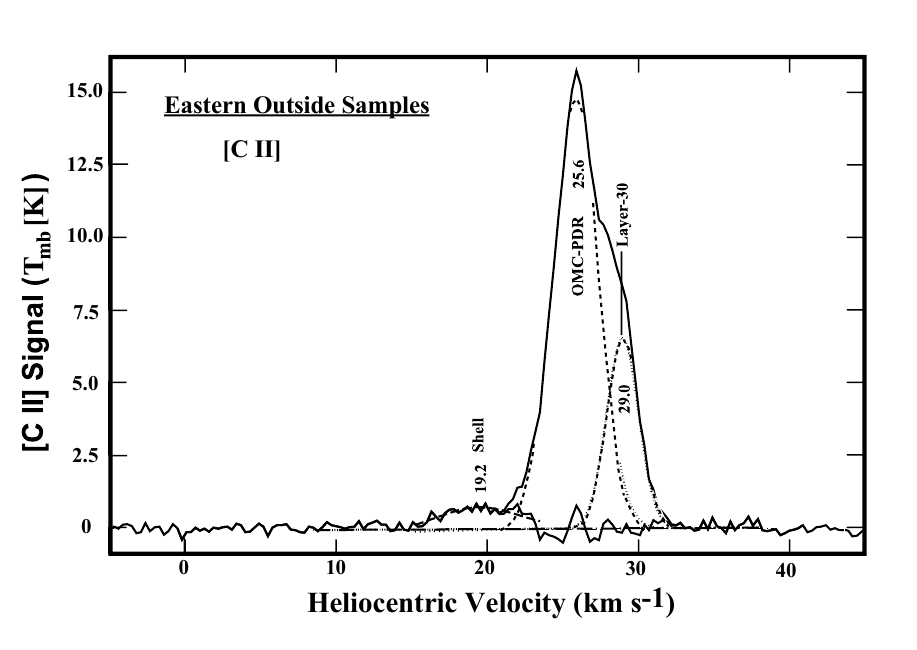}
\caption{This is an average  of  \Cii\ samples in our Outer Border group (i, ii, I, L-East, K-East, J) shown in Figure~\ref{fig:EONsamples}, all lying east of the eastern rim boundary of the EON. The identification of the components are discussed in Section~\ref{sec:Outside}. The lower heavy lines are the remaining signal after the 25.6 \kms\ and 28.6 \kms\ components have been subtracted, making the low velocity components clearer.
\label{fig:Outside}}
\end{figure*}

In order to characterize the \Cii\ emitting material lying in the Outer Border, we chose six samples to the east or clearly outside to the south of the N-S Rim (I from the Southern-Profile, One and Two from the SE-Profile, and Samples L-East, K-East, and J). The averaged spectra are shown in Figure~\ref{fig:Outside} and the components are discussed below.

 The 29.0$\pm$0.3 \kms\ component is present in all of the samples and is on the low end of those we've assigned to Layer-30.
 
The strongest component at  25.6$\pm$0.5 \kms\ is also seen in all the samples. Comparison of our outside samples with the PV-Cuts presented in this study and in \citet{pabst20} shows that they are in the enhanced PDR emission lying near (but not at) the closest approach of the Shell velocity to the PDR, as discussed in Section~\ref{sec:structure}. They are not part of the Shell.
 
 The 19.2 \kms\ component shown is from two samples (20.5 \kms\ in sample I and 18.6 \kms\ in sample J) and the PV-Cuts show that they are parts of the Shell. They are similar to the 20.0$\pm$0.8 \kms\ component in \hr.

\begin{table}
\caption{Inside and Outside Velocity Systems*}
\label{tab:COplusCII}
\begin{tabular}{lllll}
\hline
\hline
\colhead{Locations} &
\colhead{Pabst(2022) \Cii} &
\colhead{This Study \Cii} &
\colhead{Source}&
\colhead{Pabst(2022) CO}\\
\hline       
{\bf Inside}**  & (...)                         & 30.3$\pm$0.6         &Layer-30       & 29.0$\pm$0.4\\   
(...)              & {\bf 27.6$\pm$1.0} & {\bf 26.6$\pm$0.6 } &OMC-PDR    &{\bf 26.9$\pm$0.7}\\
(...)              &  20.1$\pm$1.0        & 18.4$\pm$1.9         &Shell             & (...)\\
(...)              &  14.8$\pm$3           &    (...)                        &(...)                   &(...)\\
{\bf Outside***}      &29.1$\pm$0.7  & 29.0$\pm$0.3          & Layer-30                &{\bf 29.3$\pm$1.0}\\
(...)                 &{\bf 25.7$\pm$0.4}& {\bf 25.6$\pm$0.5}  &OMC-PDR    & 26.0$\pm$0.5\\
(...)              &21.9$\pm$0.4      & 19.2$\pm$1.5           &Shell                 & (...) \\
 \hline
\end{tabular}\\

~*All velocities are \kms\ Heliocentric. The strongest component is in {\bf boldface}.

**\citet{pabst22} samples Green 3--6, White 1, 3, and 5; This study samples Huygens Region, SE-Profile 4--11, Middle-Profile II--VII, Southern-Profile iii--vii. 

***\citet{pabst22} samples Green 1--2, White 2, 4, and 6; This study samples, I, L-East, K-East, J, Southern-Profile i--ii. 

\end{table}

A useful similar analysis can be done using results from \citet{pabst22}. In that study twelve circular samples of 40\arcsec\ radius were made across \hr , the EON, 
and the Outer Border in both \Cii\ and CO(2-1) emission. We have grouped their samples into those inside and outside the dashed line in Figure~\ref{fig:EONsamples}. The averaged results are presented in Table~\ref{tab:COplusCII}, in addition to the \Cii\ results in this study. We see that the \Cii\ results are very similar, especially considering that the sample sizes were not exactly the same size. 

For the CO samples, the Inside average of 26.9$\pm$0.7 \kms\ is very similar to 27.3 \kms\ that we have being using from CO emission for the OMC-PDR. However, the strength of the Layer-30 component at 29.3$\pm$1.0 \kms\ is unexpected and is discussed in Section~\ref{sec:30}.

\section{DISCUSSION}
\label{sec:Disc}

\subsection{The Layer-30 Component}
\label{sec:30}

The velocity component we call Layer-30 was first recognized in the \hi\ observations of \citet{vdw13} where it is seen along the Orion Bar, but also in extended regions as far south as 500\arcsec . The Orion Bar features certainly are associated with that escarpment, but the origin of the extended region emission and absorption is not clear. We see it in \hr\ in emission (31.5 \kms , Figure~\ref{fig:CIIandLine} and 30.6 \kms , Figure~\ref{fig:HRspec}), the SE-4 sample in emission (28.5 \kms\ and 31.1 \kms ,Figure~\ref{fig:SE4}), the Central Northern Profile in emission ($\approxeq$28--33 \kms ) and absorption ($\approxeq$28--32 \kms ) (Figure~\ref{fig:HReastwest}), and in the Dark Bay in absorption (32.1\kms , Figure~\ref{fig:DarkBay}).  

For the first time, we also see \Cii\ velocity components. We see the Layer-30 \Cii\ components in the EW Profile (Figure~\ref{fig:EWProfiles}), the NS Profile (Figure~\ref{fig:NSProfiles}), in the west of the South Profile (Figure~\ref{fig:SouthernProfile}), in the Middle Profile (Figure~\ref{fig:MiddleProfile}), the SE-4 sample (Figure~\ref{fig:SE4}), the Central Huygens Region Profile (Figure~\ref{fig:HReastwest}), the total Huygens Region spectrum (Figure~\ref{fig:HRspec}), the Dark Bay (Figure~\ref{fig:DarkBay}), and the Outside Samples (Figure~\ref{fig:Outside}).
Since they always appear on the difficult to deconvolve red shoulder of the OMC-PDR component they may be even more ubiquitous. 

A quantitative analysis of the Layer-30 is probably done best using the spectra of the full Huygens region, where it is seen
in emission in both \hi\ at 30.6 \kms\ and \Cii\ at 31.0 \kms . This region is known to show \Nai\ and \Caii\ absorption lines at 31.3 \kms\ and 30.8 \kms\  \citep{ode93} and these may be related to the Layer-30 material. If so, then that layer must lie
in the foreground of the ONC.

\subsection{Structure of the Outer Shell}
\label{sec:structure}

\subsubsection{Radial Velocity Convergence}
The expectation of the pattern expected in a limb-brightened shell is that the radial velocity would increase abruptly as the line-of-sight approaches
the apparent maximum size of the Shell. This is not what we see in the profiles or the PV-Cuts. The outer Shell radial velocities increase (approach the velocity of the underlying OMC) gradually and usually have flattened before reaching the OMC velocity, rather than rapidly changing. We also see this in the PV-Cuts presented in Appendix C of \citet{pabst20}. 

\subsubsection{Limb-Brightening}
In the discovery paper of the Shell \citet{pabst19} referred to the bright outer  region (our Outer Border) as due to Limb-Brightening of the Shell, an assumption 
and nomenclature used in multiple subsequent papers \citet{pabst20,pabst22,goi20,kavak22a}. However, that interpretation deserves re-examination since
it is key to understanding the physical nature of the Shell. We have examined the profiles and PV-Cuts in this paper with this in mind. 
Since the western ends of the profiles clearly indicate that the Shell never quite reaches the OMC velocity, we limit our discussion to the east side of the profiles.

The expectation for the brightness of a closed shell intersecting the OMC is that there would be a narrow region of increased surface brightness. Examination of our PV-Cuts shows this is not the case. On the east end of the Northern Profile PV-Cut we see that the brightness enhancement peaks before the distance of convergence. A similar thing is seen in the SE-Profile, Middle-Profile, and Southern-Profile PV-Cuts. There are enhancements at nearly constant velocities in the region outside of 500\arcsec\ east, but these are not linked to 
the Shell components. Definitely the background Outer Region is enhanced \Cii\ brightness, but not because of Shell limb-brightening.  

\subsubsection{An Altered Radiation Field Illuminates the Outer Region}
\label{sec:altered}

The enhancement in the Outer Region \Cii\ brightness is probably due to the very different radiation field that it sees, rather than limb-brightening of the Shell.

It is well established that the observed continuum in \hr\ is much stronger than expected \citep{bal91,ode10}.
The most common measure of this is the Equivalent Width (EW), typically measured at the wavelength of \Hb , with the EW being the ratio of the emission line signal and the underlying continuum, usually expressed in $\rm \AA$ . It is the interval of the continuum required to equal the flux from the central emission line. A lower EW means a relatively stronger continuum. For an ionized gas at 9000 K and density of 6000 \cmq\ the expected EW from atomic processes is 1800 $\rm \AA$ and for a density of 
200 \cmq\ 1600 $\rm \AA$ \citep{ode10}. The low EW values mean that scattered starlight dominates the continuum. These low EW values become a measure of hydrogen ionizing photons (EUV) to non-ionizing photons, that includes the FUV that determines the emissivity of \Cii . As the FUV becomes relatively stronger, one expects the EW to become smaller and the relative emissivity of \Cii\ to increase.
 
From the  extensive map of the EW presented in \citet{ode10} we find the EW's for the Inside Samples of \citet{pabst22} are 320$\pm$80 $\rm \AA$. This indicates that an increased FUV/EUV ratio is important even within the EON. This is also demonstrated \citep{ode10} by the EW decreasing from \hr\ into the southern EON,  reaching values of about 150 $\rm \AA$. 

In \citet{ode09} it was established that the conditions of illumination change dramatically as the line-of-sight transitions from the EON into what we now call the Outer Region. This was done by comparing the derived emission measure in \Hb\ and 327.5 MHz thermal radio continuum. It was found that in the Outer Region 
much of the observed \Hb\ was due to local scattering of Huygens Region \Hb\ emission. The low EW values seen in the Outer Region would then be 
due to decreased local emission of \Hb\ owing to illumination by a modified radiation field with a low EUV/FUV ratio.

Using the Outer Border sample regions in \citet{pabst22} we find an average EW of 170$\pm40$ $\rm \AA$, indicating that the Outer Border sees a very different radiation field than the Inner Samples and one that is similar to those in the southern part of the EON. This is probably because the N-S Rim is a boundary where the local material transitions to seeing less EUV radiation. This means that the emissivity of \Cii\ should be enhanced there, explaining most
of the brightness of the Outer Border. 

This filtering out of EUV radiation due to trace neutral hydrogen will be most important in directions near the MIF, whereas the foreground material forming the \Cii\ Shell would directly see radiation from the Trapezium stars that has not been modified by filtering out the EUV radiation.

\subsubsection{The \Cii\ Shell and the \hi\ Veil-B Component Form the Outer Shell}
\label{sec:Shell-B}

We have seen that in the region of overlap of the \hi\ and \Cii\ lines that one velocity component agrees in both emitters (Figure~\ref{fig:SE4},\ref{fig:HReastwest},\ref{fig:HRspec},\ref{fig:DarkBay}). That component was called Veil-B in \hi\ studies and the Shell in \Cii\ studies. The coincidence of velocities are strong evidence that this \hi\ and \Cii\ emission come from the same physical feature, probably from slightly different regions within that feature. We now will refer to this feature as the Outer Shell.  

\subsubsection{Our Model for the Outer Shell}

\begin{figure*}
\epsscale{0.8}
\plotone{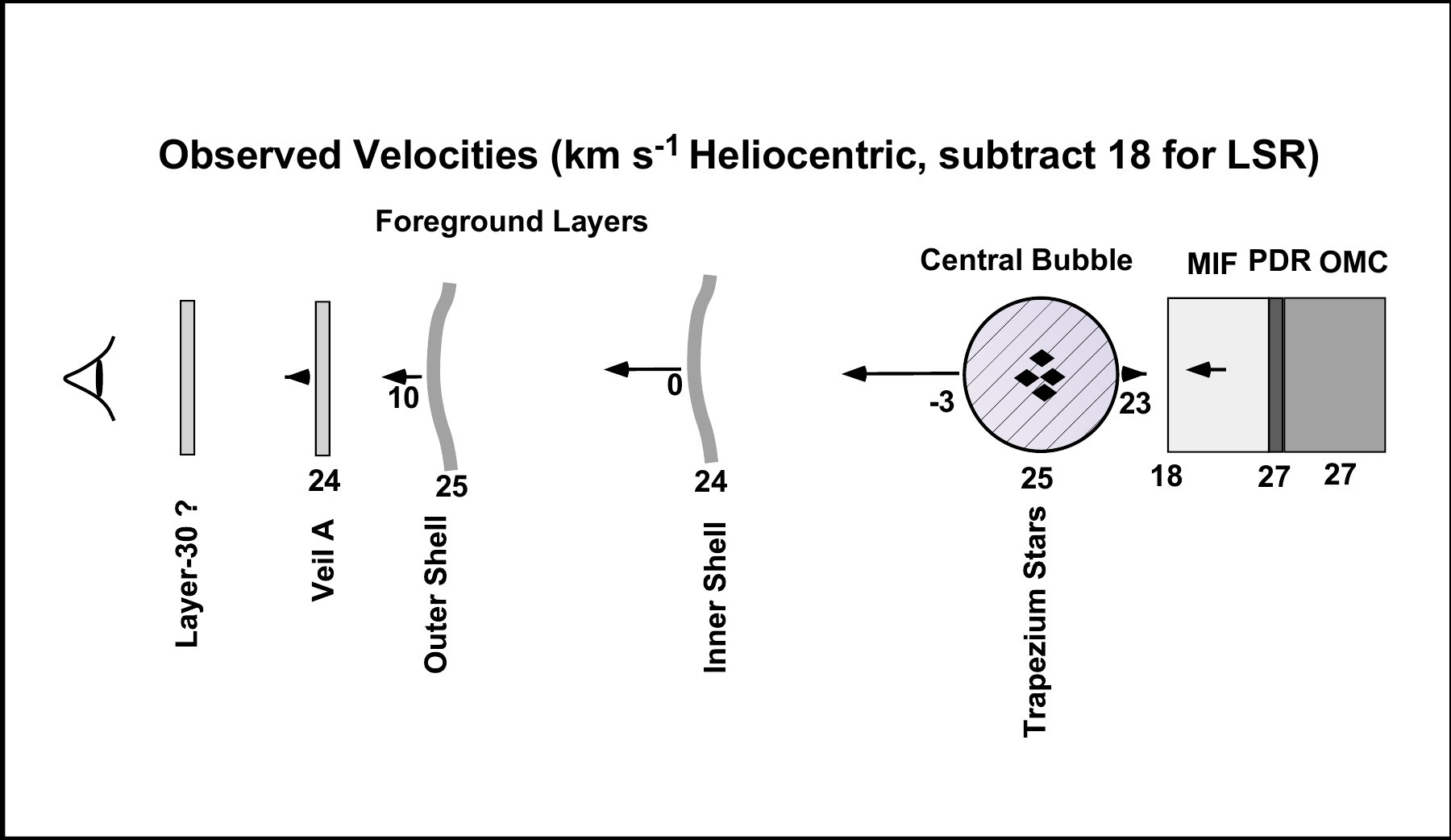}
\caption{This model depicts in cartoon fashion the multiple features seen in the direction of the Orion Nebula and it's associated Extended Orion Nebula. Figure~\ref{fig:Model} is an updated version of Figure 8 in \citet{ode20} and is contrast with the simplified two dimensional drawing in \citet{pabst19}. 
There are important differences that we note here from right to left. The High Ionization Central Bubble seen in the inner Huygens Region \cite{ode18} freely expands towards the observer, but is slowed on the far side by outflowing photo-ionized gas in the Main Ionization Front \citep{ode20}. There is a ionized layer designated as the Inner Shell. It is a bulge in a foreground layer of material at about 24 \kms\ and has a maximum
expansion velocity of 27 \kms . Outside of this is the Outer Shell, which is again a wind-driven bulge in a foreground layer at about 25 \kms\ and has a maximum expansion velocity of 15 \kms. It is composed of the \Cii -Shell detailed in \citet{pabst20} and the \hi\ Veil-B material announced in \citet{vdw89}. The \hi\ Veil-A feature shows no \Cii\ component. The Layer-30 component is present in both \hi\ and \Cii . It lies in the foreground if the ionic absorption lines seen in the Trapezium stars at the same velocity are due to it. If those lines are not, this material could lie in the background.
\label{fig:Model}}
\end{figure*}

Prior to the discovery of the \Cii\ component, the Outer Shell was modeled quantitatively. In \citet{abel16}, it was called Veil-B. In \citet{abel19} it was called Component III. Abel et al (2016) computed the physical conditions using radio, optical, and UV absorption data. The Model derived in the present study is shown in Figure~\ref{fig:Model}.

  The [C II] observation provides an added constraint to this model.  The observed [C II] intensity is 5.12$\times$10$\rm^{-4}$ erg cm$\rm^{-2}$ s$\rm^{-1}$ sr$\rm^{-1}$, with an estimated range of +/- 30$\%$. Taking the best fit model from Table 5 of Abel et al. (2016), we calculate a predicted [C II] intensity of 8$\times$10$\rm^{-5}$ erg cm$\rm^{-2}$ s$\rm^{-1}$ sr$\rm^{-1}$, a factor of six lower than observed.  This discrepancy could be due to several factors.  First, it could be the C/H ratio. \citet{sofia04} found a C/H ratio of (137$\pm$56)$\times$10$\rm ^{-6}$.  Our model uses this central value for the C/H abundance ratio.  Decreasing the C/H ratio would decrease the predicted [C II] emission.  Second, it could be the assumed HI column density. \citet{cart01} found the total \hi\ column density along the line of sight towards the Trapezium to be  (4.8 $\pm$1.1)x10$\rm^{21}$ cm$\rm^{-2}$.    
STIS observations of elements in ionization stages associated with HI gas show about 2/3 of their total column density correlating with the same velocity as Component III. We adopt that ratio. Using the lower range for the HI column density from \citet{cart01} would decrease the predicted [C II] emission by decreasing the C+ column density.  Finally, moving component III farther from the Trapezium would decrease [C II] emission by decreasing incident radiation G0 (e.g \citet{kaufman99}).  While each of these scenarios (along with the possibility of the observed [C II] emission being as much as 30$\%$ less than the central observed value) would move the model more into agreement with the observed [C II] emission in this paper. A revised model would have to also account for the pre-existing observations from STIS, in particular the molecular hydrogen absorption spectrum presented in \citet{abel16,abel19}.  This revised model, along with a model of Layer 30, will be the focus of a future work.

\subsubsection{A Recent Model for the Outer Shell}
\label{sec:shell}
In \citet{pabst19} a model is presented where the Outer Shell component is an expanding closed hemisphere that touches the OMC in a limb-brightened outer shell, with the Outer Shell driven by strong stellar winds from \tC . The picture derived in the present study is different in important ways. This model was based on \Cii\ observations alone and did not accommodate observations of the blue \Cii\ emission, and 
\hi\ and visual and UV ionization gas observations.

Our model for the Outer Shell is that it is a  part of a physical feature in foreground material at a velocity slightly smaller than the OMC-PDR that lies between the inner high ionization shocked gas and the outer \hi\ Veil-A material. The region overlying \hr\ and the EON has been accelerated towards the observer, forming a large bulge that in cross-section
resembles a shell. This bulge is surrounded by material having the undisturbed velocity of about 26 \kms.

\subsubsection{Modeling the Inner Shell}
\label{sec:BlueShells}

There exists a  second shell structure (the Inner Shell) that reaches to about 0 \kms\ (an expansion
velocity of 27 \kms ), that is seen in the the earlier PV-Cuts, but not recognized. There is marginal evidence for another shell extending to less than -20 \kms. 

The higher radial velocity components are shown in Figures \ref{fig:CIIandLine}, \ref{fig:EWProfiles}, \ref{fig:NSProfiles}, \ref{fig:MiddleProfile}, \ref{fig:SE4}, \ref{fig:HReastwest}, \ref{fig:HRspec}, and \ref{fig:DarkBay}. The correlation of \hi\ and \Cii\ features is not simple, with different lines being stronger on a case by case basis. Sometimes the \hi\ features are well defined and in other cases present only as spectrally unresolved emission.

The visual \sii\ doublet components of what we now recognize as part of the Inner Shell were presented in \citet{gar07}. Since then multiple emission
and absorption lines have been added (c.f. Table 2 of \citet{abel19} and with the present study we see that these are part of an expanding shell (Figure~\ref{fig:EWProfiles}), rather than a simple blue-shifted physical feature. \citet{abel19} modeled the available blue-shifted observations, assigning most of them to a Component I and the few, more red-shifted \oi\ (11 \kms) observations to a Component II. The new \hi\ and \Cii\ should be added to the brew used in building a model of this inner region. The well defined \Cii\ data should better define any ionization transition zone than the difficult-to-observe optical \oi\ observations that compete with 
night sky emission at the same wavelength. 

\subsubsection{Driving Forces}
\label{sec:forces}

In \citet{pabst19} the hemispherical shell is explained as the result of a fast stellar wind from \tC, outside of the inner High Ionization sphere of shocked region established from optical observations \citet{ode18}.  This same wind could account for a hemispherical bulge in a foreground layer. However, the presence of the higher expansion velocity Inner Shell calls for re-examination of this simple model as an additional shell demands an additional driving force. The simplest, but ad hoc explanation is that \tC\ or another hot inner-cluster star underwent a more recent fast stellar wind phase.  We know that this additional wind must be more recent as \citet{abel19} has established that the Inner Shell is the closest component to \tC .

\subsection{Related Studies}
\label{sec:others}

In a study of a velocity resolved map of CO emission, \citet{goi20} determined that there was no extended emission that would be coming from the Outer Shell. This indicates that FUV from the Trapezium has photo-dissociated any original molecular component in the Outer Shell. However, they did find 10 compact CO sources at velocities corresponding to the expanding \Cii\ Shell in their locations (shown in Figure~\ref{fig:Northernsamples} and Figure~\ref{fig:EONsamples}). They are usually of substellar mass. They
do not address how these objects formed within the shell can continue to share the Shell's expansion even when the driving  force per unit cross-section will have dropped with their creation. 

\citet{kavak22b} has searched the same \Cii\ and \hi\ data bases, supplemented by observations in CO, 8 \micron\ PAH, and 70 \micron\  thermal emission, looking for structures indicating collimated stellar outflow interacting with the Shell. Six objects were found in \Cii , none having CO
or \hi\ counterparts.  Adopting their name Dents, we show their positions in Figure~\ref{fig:Northernsamples} and a characteristic Dent (4) appears in the lower 
panel of Figure~\ref{fig:PVnorthern}. Their basic interpretation of shocks formed by collimated stellar outflows as they pass through the Outer Shell seems correct. Dents 1 through 4 are the best defined, all of which start with the 20 \kms\ Shell velocity and extend to the blue by 6 to 13 \kms (into the velocity range of the Inner Shell). However, if the higher expansion velocity Inner Shell is closer to \tC\ than the Outer Shell, that interpretation is problematic.

\section{CONCLUSIONS}
\label{sec:end}

$\bullet$ The coincidence of velocities of the \Cii\ Shell and the \hi\ Veil-B in the Huygens Region argues that they are the same physical feature and should be designated as the Outer Shell. This has a maximum expansion velocity of 15 \kms .

$\bullet$  The incomplete merger of the \Cii\ Shell and OMC-PDR velocities argues that the former is actually a concavity in a foreground layer of gas.

$\bullet$ A second expanding apparent half shell seen in all our data sets is present, with a maximum expansion velocity of 27 \kms. 
Its concentration to \hr\ argues that it is the result of a strong wind more recent than that producing the slower outer concavity. It is designated as the Inner Shell.

$\bullet$ The Outer Region surrounding \hr\ and the EON is not bright due to limb-brightening of the \Cii\ Shell, rather, it is primarily due to 
emission from a PDR illuminated by Trapezium stars that have had most of their EUV filtered out by intervening residual neutral hydrogen.

$\bullet$ The embedded Ori-S molecular cloud shows a PDR on the observer's and farther sides, in addition to the OMC-PDR lying beyond the cloud. We also see Ori-S PDR material blue-shifted by the outflows feeding the major shocks in \hr .

$\bullet$ \Cii\ samples of HH~269 at and to the west from the Ori-S molecular cloud show the same velocity components as that cloud, 
indicating that the cloud is simply a dense region within an extended structure. Only in this region do we see the \hi\ Veil-C and van der Werf's 
(2013) Shock-C components. 

$\bullet$ The Dark Bay shows all the components of \hr\ including the OMC-PDR, but the velocities are redshifted by 1.9$\pm$0.4 \kms , indicating that the northeast section of the OMC and its foreground layers are moving redshifted by this amount.

$\bullet$ Layer-30 features are seen across \hr , the Dark Bay, and much of the northern EON. If this is the source of \Nai\ and \Caii\ absorption lines at the same velocities, then it must lie in the foreground of the OMC.

\section*{acknowledgements}
We wish to acknowledge the cooperation of Cornelia H. M. Pabst, \"Umit Kavak,  and Paul van der Werf of the Leiden Observatory, Leiden University, the Netherlands for providing us the \Cii\ and \hi\ data used in this study. Mark Manner of Nashville, Tennessee created and permitted use of the images used in Figure~\ref{fig:Northernsamples} and Figure~\ref{fig:EONsamples}. 

\facilities{VLA},{SOPHIA},{HST},{CTIO},{SPM},{KPNO}

\appendix

\section{Creation of the Full Orion Nebula Image}
\label{sec:Mark}

The image of M42 was acquired by Mark Manner, on the Cumberland Plateau in middle Tennessee using a 130 mm aperture apochromatic refractor operating at f/5.  The imaging camera contained a Sony back-illuminated IMX571 CMOS sensor. The sensor contained 6248x4176 3.76-micron pixels, with a Bayer RGB filter array incorporated on top of the sensor to allow preparation of an RGB image. The scale was 1\farcs18 /pixel. Two hundred and fourteen 60 second unguided individual exposures were taken with the camera cooled to 0\arcdeg\ C.  

Post observation image processing (dark and flat calibration, normalization, registration and integration) was performed using Pleiades Astrophoto's  Pixinsight software. The resulting master integrated linear image had a permanent histogram transformation applied in Pixinsight to reveal both light and dark regions of the nebula, with a final contrast enhancement performed in Adobe Photoshop. The mid-point time of the data acquisition was UTC 2024-10-24T07:53:19. The image is available at https://astrob.in/kp8lch/G/ .

\end{document}